\documentclass[longauth]{aa}
\usepackage{euclid}
\usepackage[varg]{txfonts}
\usepackage[]{natbib}
\usepackage[labelfont=bf]{caption}
\usepackage{adjustbox}
\usepackage[super]{nth}
\usepackage[]{siunitx}
\usepackage{subfig}
\usepackage{mathtools}
\usepackage{amsmath}
\usepackage{bm}
\usepackage{amssymb}
\usepackage{graphicx}
\usepackage{enumerate}

\newcommand{\sn}{{\rm S/N}\xspace}
\newcommand{\euc}{\textit{Euclid}\xspace}
\newcommand{\uj}{ULAS J1120+0641\xspace}

\newcommand{\lya}{Ly\(\alpha\)\xspace}

\newcommand{\pq}{\(P_{\mathrm{q}}\)\xspace}
\newcommand{\dsq}{{\rm deg}^2}
\newcommand{\chisq}{\chi^2}
\newcommand{\given}{\,\middle\vert\,}

\newcommand{\zf}{z_{\mathrm f}}
\newcommand{\Civ}{{\rm C}\,\textsc{iv}\xspace}
\newcommand{\Mstar}{\(M_*\)\xspace}
\newcommand{\myrange}[2]{{#1}\,\textrm{--}\,{#2}\xspace}
\usepackage{dcolumn}
\makeatletter
\newcommand*{\smallrel}[2][.8]{%
	\mathrel{\mathpalette{\smallrel@{#1}}{#2}}%
}
\newcommand*{\smallrel@}[3]{%
	\sbox0{$#2\vcenter{}$}%
	\dimen@=\ht0 %
	\raise\dimen@\hbox{%
		\scalebox{#1}{%
			\raise-\dimen@\hbox{$#2#3\m@th$}%
		}%
	}%
}
\makeatother

\begin{document}
	\urlstyle{same}
	\title{\euc preparation: V. Predicted yield of redshift \(7<z<9\) quasars from the wide survey}
\author{\Euclid Collaboration, R.~Barnett$^{1}$, S.J.~Warren$^{1}$, D.J.~Mortlock$^{1,2,3}$, J.-G.~Cuby$^{4}$, C.~Conselice$^{5}$, P.C.~Hewett$^{6}$, C.J.~Willott$^{7}$, N.~Auricchio$^{8}$, A.~Balaguera-Antolínez$^{9}$, M.~Baldi$^{8,10,11}$
% ...
}

\institute{$^{1}$ Astrophysics Group, Blackett Laboratory, Imperial College London, London SW7 2AZ, UK\\
	$^{2}$ Department of Astronomy, Stockholm University, Albanova, SE-10691 Stockholm, Sweden\\
	$^{3}$ Department of Mathematics, Imperial College London, London SW7 2AZ, UK\\
	$^{4}$ Aix-Marseille Univ, CNRS, CNES, LAM, Marseille, France\\
	$^{5}$ University of Nottingham, University Park, Nottingham NG7 2RD, UK\\
	$^{6}$ Institute of Astronomy, University of Cambridge, Madingley Road, Cambridge CB3 0HA, UK\\
	$^{7}$ NRC Herzberg, 5071 West Saanich Rd, Victoria, BC V9E 2E7, Canada\\
	$^{8}$ INAF-Osservatorio di Astrofisica e Scienza dello Spazio di Bologna, Via Piero Gobetti 93/3, I-40129 Bologna, Italy\\
	$^{9}$ Instituto de Astrof\'{i}sica de Canarias. Calle V\'{i}a L\`{a}ctea s/n, 38204, San Crist\'{o}bal de la Laguna, Tenerife, Spain\\
	$^{10}$ Dipartimento di Fisica e Astronomia, Universit\'a di Bologna, Via Gobetti 93/2, I-40129 Bologna, Italy\\
	$^{11}$ INFN-Sezione di Bologna, Viale Berti Pichat 6/2, I-40127 Bologna, Italy\\
	$^{12}$ Max Planck Institute for Extraterrestrial Physics, Giessenbachstr. 1, D-85748 Garching, Germany\\
	$^{13}$ Universit\"ats-Sternwarte M\"unchen, Fakult\"at f\"ur Physik, Ludwig-Maximilians-Universit\"at M\"unchen, Scheinerstrasse 1, 81679 M\"unchen, Germany\\
	$^{14}$ INAF-Osservatorio Astronomico di Trieste, Via G. B. Tiepolo 11, I-34131 Trieste, Italy\\
	$^{15}$ INAF-Osservatorio Astrofisico di Torino, Via Osservatorio 20, I-10025 Pino Torinese (TO), Italy\\
	$^{16}$ Department of Astronomy, University of Geneva, ch. d'\'Ecogia 16, CH-1290 Versoix, Switzerland\\
	$^{17}$ INFN-Sezione di Roma Tre, Via della Vasca Navale 84, I-00146, Roma, Italy\\
	$^{18}$ Department of Mathematics and Physics, Roma Tre University, Via della Vasca Navale 84, I-00146 Rome, Italy\\
	$^{19}$ INAF-Osservatorio Astronomico di Roma, Via Frascati 33, I-00078 Monteporzio Catone, Italy\\
	$^{20}$ INAF-Osservatorio Astronomico di Capodimonte, Via Moiariello 16, I-80131 Napoli, Italy\\
	$^{21}$ Instituto de Astrof\'isica e Ci\^encias do Espa\c{c}o, Universidade do Porto, CAUP, Rua das Estrelas, PT4150-762 Porto, Portugal\\
	$^{22}$ Dipartimento di Fisica e Scienze della Terra, Universit\'a degli Studi di Ferrara, Via Giuseppe Saragat 1, I-44122 Ferrara, Italy\\
	$^{23}$ INAF, Istituto di Radioastronomia, Via Piero Gobetti 101, I-40129 Bologna, Italy\\
	$^{24}$ INFN-Sezione di Torino, Via P. Giuria 1, I-10125 Torino, Italy\\
	$^{25}$ Dipartimento di Fisica, Universit\'a degli Studi di Torino, Via P. Giuria 1, I-10125 Torino, Italy\\
	$^{26}$ INFN-Sezione di Milano, Via Celoria 16, I-20133 Milan, Italy\\
	$^{27}$ INAF-IASF Milano, Via Alfonso Corti 12, I-20133 Milano, Italy\\
	$^{28}$ Institut de F\'isica d’Altes Energies IFAE, 08193 Bellaterra, Barcelona, Spain\\
	$^{29}$ Instituto de Astrof\'isica e Ci\^encias do Espa\c{c}o, Faculdade de Ci\^encias, Universidade de Lisboa, Tapada da Ajuda, PT-1349-018 Lisboa, Portugal\\
	$^{30}$ Institute of Space Sciences (ICE, CSIC), Campus UAB, Carrer de Can Magrans, s/n, 08193 Barcelona, Spain\\
	$^{31}$ Institut d’Estudis Espacials de Catalunya (IEEC), 08034 Barcelona, Spain\\
	$^{32}$ Department of Physics "E. Pancini", University Federico II, Via Cinthia 6, I-80126, Napoli, Italy\\
	$^{33}$ INFN section of Naples, Via Cinthia 6, I-80126, Napoli, Italy\\
	$^{34}$ INAF - Osservatorio Astrofisico di Arcetri, Largo E. Fermi 5, I-50125, Firenze, Italy\\
	$^{35}$ Centre National d'Etudes Spatiales, Toulouse, France\\
	$^{36}$ Institute for Astronomy, University of Edinburgh, Royal Observatory, Blackford Hill, Edinburgh EH9 3HJ, UK\\
	$^{37}$ ESAC/ESA, Camino Bajo del Castillo, s/n., Urb. Villafranca del Castillo, 28692 Villanueva de la Ca\~nada, Madrid, Spain\\
	$^{38}$ Universit\'e de Lyon, F-69622, Lyon, France ; Universit\'e de Lyon 1, Villeurbanne; CNRS/IN2P3, Institut de Physique Nucl\'eaire de Lyon, France\\
	$^{39}$ Mullard Space Science Laboratory, University College London, Holmbury St Mary, Dorking, Surrey RH5 6NT, UK\\
	$^{40}$ Departamento de F\'isica, Faculdade de Ci\^encias, Universidade de Lisboa, Edif\'icio C8, Campo Grande, PT1749-016 Lisboa, Portugal\\
	$^{41}$ Instituto de Astrof\'isica e Ci\^encias do Espa\c{c}o, Faculdade de Ci\^encias, Universidade de Lisboa, Campo Grande, PT-1749-016 Lisboa, Portugal\\
	$^{42}$ Department of Physics, Oxford University, Keble Road, Oxford OX1 3RH, UK\\
	$^{43}$ INFN-Padova, Via Marzolo 8, I-35131 Padova, Italy\\
	$^{44}$ Institut de Physique Nucl\'eaire de Lyon, 4, rue Enrico Fermi, 69622, Villeurbanne cedex, France\\
	$^{45}$ Aix-Marseille Univ, CNRS/IN2P3, CPPM, Marseille, France\\
	$^{46}$ AIM, CEA, CNRS, Universit\'{e} Paris-Saclay, Universit\'{e} Paris Diderot, Sorbonne Paris Cit\'{e}, F-91191 Gif-sur-Yvette, France\\
	$^{47}$ Institut de Ciencies de l'Espai (IEEC-CSIC), Campus UAB, Carrer de Can Magrans, s/n Cerdanyola del Vall\'es, 08193 Barcelona, Spain\\
	$^{48}$ Centre for Extragalactic Astronomy, Department of Physics, Durham University, South Road, Durham, DH1 3LE, UK\\
	$^{49}$ Leiden Observatory, Leiden University, Niels Bohrweg 2, 2333 CA Leiden, The Netherlands\\
	$^{50}$ von Hoerner \& Sulger GmbH, Schlo{\ss}Platz 8, D-68723 Schwetzingen, Germany\\
	$^{51}$ Max-Planck-Institut f\"ur Astronomie, K\"onigstuhl 17, D-69117 Heidelberg, Germany\\
	$^{52}$ Institut d'Astrophysique de Paris, 98bis Boulevard Arago, F-75014, Paris, France\\
	$^{53}$ Department of Physics and Helsinki Institute of Physics, Gustaf H\"allstr\"omin katu 2, 00014 University of Helsinki, Finland\\
	$^{54}$ Universit\'e de Gen\`eve, D\'epartement de Physique Th\'eorique and Centre for Astroparticle Physics, 24 quai Ernest-Ansermet, CH-1211 Gen\`eve 4, Switzerland\\
	$^{55}$ European Space Agency/ESTEC, Keplerlaan 1, 2201 AZ Noordwijk, The Netherlands\\
	$^{56}$ Institute of Theoretical Astrophysics, University of Oslo, P.O. Box 1029 Blindern, N-0315 Oslo, Norway\\
	$^{57}$ Institute of Space Sciences (IEEC-CSIC), c/Can Magrans s/n, 08193 Cerdanyola del Vall\'es, Barcelona, Spain\\
	$^{58}$ Argelander-Institut f\"ur Astronomie, Universit\"at Bonn, Auf dem H\"ugel 71, 53121 Bonn, Germany\\
	$^{59}$ Institute for Computational Cosmology, Department of Physics, Durham University, South Road, Durham, DH1 3LE, UK\\
	$^{60}$ INAF-Osservatorio Astronomico di Padova, Via dell'Osservatorio 5, I-35122 Padova, Italy\\
	$^{61}$ University of Paris Denis Diderot, University of Paris Sorbonne Cit\'e (PSC), 75205 Paris Cedex 13, France\\
	$^{62}$ Sorbonne Universit\'e, Observatoire de Paris, Universit\'e PSL, CNRS, LERMA, F-75014, Paris, France\\
	$^{63}$ CEA Saclay, DFR/IRFU, Service d'Astrophysique, Bat. 709, 91191 Gif-sur-Yvette, France\\
	$^{64}$ INAF-IASF Bologna, Via Piero Gobetti 101, I-40129 Bologna, Italy\\
	$^{65}$ Observatoire de Sauverny, Ecole Polytechnique F\'ed\'erale de Lau- sanne, CH-1290 Versoix, Switzerland\\
	$^{66}$ INFN-Bologna, Via Irnerio 46, I-40126 Bologna, Italy\\
	$^{67}$ Kapteyn Astronomical Institute, University of Groningen, PO Box 800, 9700 AV Groningen, The Netherlands\\
	$^{68}$ Perimeter Institute for Theoretical Physics, Waterloo, Ontario N2L 2Y5, Canada\\
	$^{69}$ Department of Physics and Astronomy, University of Waterloo, Waterloo, Ontario N2L 3G1, Canada\\
	$^{70}$ Centre for Astrophysics, University of Waterloo, Waterloo, Ontario N2L 3G1, Canada\\
	$^{71}$ Space Science Data Center, Italian Space Agency, via del Politecnico snc, 00133 Roma, Italy\\
	$^{72}$ Jet Propulsion Laboratory, California Institute of Technology, 4800 Oak Grove Drive, Pasadena, CA, 91109, USA\\
	$^{73}$ Departamento de F\'isica, FCFM, Universidad de Chile, Blanco Encalada 2008, Santiago, Chile\\
	$^{74}$ I.N.F.N.-Sezione di Roma Piazzale Aldo Moro, 2 - c/o Dipartimento di Fisica, Edificio G. Marconi, I-00185 Roma, Italy\\
	$^{75}$ Centro de Investigaciones Energ\'eticas, Medioambientales y Tecnol\'ogicas (CIEMAT), Avenida Complutense 40, 28040 Madrid, Spain\\
	$^{76}$ Phys. Dep. Universit\`a di Milano-Bicocca, Piazza della scienza 3, Milano, Italy\\
	$^{77}$ INFN-Sezione di Milano-Bicocca, Piazza della Scienza 3, I-20126 Milan, Italy\\
	$^{78}$ Universidad Polit\'ecnica de Cartagena, Departamento de Electr\'onica y Tecnolog\'ia de Computadoras, 30202, Cartagena, Spain\\
	$^{79}$ Departamento F\'isica Aplicada, Universidad Polit\'ecnica de Cartagena, Campus Muralla del Mar, 30202 Cartagena, Murcia, Spain\\
	$^{80}$ Infrared Processing and Analysis Center, California Institute of Technology, Pasadena, CA 91125, USA\\
	\email{rhys.barnett09@imperial.ac.uk}}
	\date{Received <> / Accepted <>}
	
	\abstract{We provide predictions of the yield of \(7<z<9\) quasars from the \euc wide survey,
	updating the calculation presented in the \euc Red Book in several ways.
	We account for revisions to the \euc near-infrared filter wavelengths; 
	we adopt steeper rates of decline of the quasar luminosity function 
	(QLF; \(\Phi\)) with redshift, $\Phi\propto10^{k(z-6)}$, $k=-0.72$,
	and a further steeper rate of decline, \(k=-0.92\); 
	we use better models of the contaminating populations (MLT dwarfs and compact early-type galaxies);
	and we make use of an improved Bayesian selection method, 
	compared to the colour cuts used for the Red Book calculation, 
	allowing the identification of fainter quasars, down to $J_{\mathrm{AB}}\sim23$. 
	Quasars at $z>8$ may be selected from \euc $OYJH$ photometry alone, 
	but selection over the redshift interval $7<z<8$ 
	is greatly improved by the addition of \(z\)-band data from, e.g., Pan-STARRS and LSST. 
	We calculate predicted quasar yields for the assumed values of the rate of decline of the QLF beyond $z=6$. 
	If the decline of the QLF accelerates beyond $z=6$, with $k=-0.92$, 
	\euc should nevertheless find over 100 quasars with $7.0<z<7.5$, 
	and \(\sim25\) quasars beyond the current record of $z=7.5$, including \(\sim8\) beyond $z=8.0$.
	The first \euc quasars at $z>7.5$ should be found in the DR1 data release, expected in 2024. 
	It will be possible to determine the bright-end slope of the QLF, $7<z<8$, $M_{1450}<-25$, 
	using 8\,m class telescopes to confirm candidates, 
	but follow-up with JWST or E-ELT will be required to measure the faint-end slope. 
	Contamination of the candidate lists is predicted to be modest even at $J_{\mathrm{AB}}\sim23$. 
	The precision with which $k$ can be determined over $7<z<8$ depends on the value of $k$, 
	but assuming $k=-0.72$ it can be measured to a $1\sigma$ uncertainty of 0.07.}

	\authorrunning{Barnett et al.}	
	\maketitle
	\section{Introduction}
	\label{sec:intro}

	High-redshift quasars can offer valuable insights into conditions in the early Universe.
	Spectra of quasars at redshifts \(z\gtrsim6\) are well established as probes of neutral hydrogen 
	in the intergalactic medium (IGM) during the later stages of the epoch of reionisation (EoR) 
	and can be used to chart the progress of this key event in cosmic history \citep[e.g.][]{Fan2006,Becker2015}. 
	High-redshift quasars are also of great interest in themselves.
	The discovery of supermassive black holes (SMBH) 
	with masses of order \(10^{\myrange{9}{10}}\,\Msolar\) at high redshift 
	\citep[e.g.][]{Mortlock2011,Wu2015,Banados2018} places strong constraints on SMBH formation within 1\,Gyr of the Big Bang. 
	The challenge posed to the standard model of SMBH formation by 
	Eddington-limited growth from stellar-mass seed black holes
	\citep[e.g.][]{Volonteri2010}, has led to investigation 
	of the formation of massive (\(M>10^4\)\,\Msolar) black-hole seeds 
	through direct collapse \citep{Bromm2003, Begelman2006,Ferrara2014,Dayal2019},
	or rapid growth via periods of super- or even hyper-Eddington 
	accretion from lower-mass seeds \citep[e.g.][]{Ohsuga2005,Inayoshi2016}.
	Additional tensions with standard SMBH growth models are implied by 
	the recent identification of young quasars (\(t < \myrange{10^4}{10^5}\)\,yr) at high redshift \citep{Eilers2017,Eilers2018}.
	These young quasars are distinguished on the basis of their small Lyman-\(\alpha\) (\lya) near zones, i.e.,
	highly ionised regions of the IGM surrounding quasars at high redshift,
	which allow enhanced flux transmission immediately bluewards of the \lya emission line,
	and before the onset of the \citet{Gunn1965} absorption trough \citep[e.g.][]{Cen2000,Bolton2011}.
	
	Around 150 quasars with redshifts \(6.0 < z < 6.5\) 
	have been discovered, mostly from the Sloan Digital Sky Survey \citep[SDSS; e.g.][]{Fan2006,Jiang2016},
	the Panoramic Survey Telescope and Rapid Response System 1 \citep[Pan-STARRS 1; e.g.][]{Banados2016},
	and the Hyper Suprime-Cam (HSC) on the Subaru telescope \citep[e.g.][]{Matsuoka2016}.
	Moreover, in the case of SDSS, rigorous analyses of completeness have allowed measurements 
	of the quasar luminosity function (QLF) to be extended to \(z=6\).
	The decline of the cumulative space density of quasars 
	brighter than absolute magnitude \(M_{1450}\) is typically parametrised as
	\begin{equation}
	\Phi\left(z,<M_{1450}\right)= \Phi\left(z_0,<M_{1450}\right)\,10^{k\left(z-z_0\right)},
	\end{equation}
	where \(z_0\) is an arbitrary anchor redshift.
	\citet{Fan2001a} found \(k = -0.47\pm0.15\)
	for bright quasars over the range \(3.5 < z < 5\).
	\citet{Fan2001b} subsequently measured the space density at \(z=6\),
	finding \(k=-0.47\) to be applicable over the whole range \(z=\myrange{3.5}{6}\).
	Such a decline has frequently been used to extrapolate 
	the measured QLF at \(z=6\) \citep{Jiang2008,Willott2010},
	e.g., to make predictions of yields of \(z>7\) quasars in other surveys.
	
	More recently, using deeper data from the SDSS Stripe 82 region,
	\citet{McGreer2013} found that \(k\) evolves over the redshift interval $4<z<6$,
	in that the number density declines less steeply at \(z<5\) (\(k>-0.47\)),
	and more steeply at \(z>5\) (\(k<-0.47\)).
	They quote \(k = -0.7\) for the redshift interval \(z = \myrange{5}{6}\).
	The most comprehensive measurement of the QLF at \(z\sim6\) has since come
	from the analysis of the complete sample of 47 SDSS quasars 
	\(5.7 < z < 6.4\) presented by \citet{Jiang2016}.
	They measured a rapid fall in quasar number density over \(z=\myrange{5}{6}\), 
	with \(k = -0.72\pm 0.11\), confirming the stronger evolution proposed by \citet{McGreer2013}.
	This has important consequences for searches for \(z>6\) quasars,
	since the yield will be considerably lower than predicted 
	by extrapolating the $z=6$ QLF using \(k=-0.47\), 
	e.g., by a factor 3 in going from $z=6$ to $z=8$.
	Indeed, given that the decline is accelerating, 
	the yield may be even lower than calculated using \(k=-0.72\).
	Very recently \citet{Wang2018a} measured
		\(k=-0.78\pm0.18\) between \(z=6\) and \(z=6.7\),
		consistent with the value measured over \(z=\myrange{5}{6}\). 
		The \citet{Wang2018a} result was published after we had completed all calculations for the current paper, 
		and so is not considered further here,
		but in any case within the quoted uncertainties it is consistent with the numbers assumed in this paper.
		
	At higher redshifts (\(z\gtrsim6.5\)) searches for quasars must be undertaken in the near-infrared (NIR), 
	as the signature \lya break shifts redwards of the optical \(z\) band. 
	The first quasar found at $z>6.5$ was the  \(z=7.08\) quasar ULAS J1120+0641 \citep{Mortlock2011},
	discovered in the UKIDSS Large Area Survey (LAS). 
	This is one of five quasars now known at $z>7$.
	Discovered more recently, ULAS J1342+0928, \(z=7.54\) \citep{Banados2018}, 
	also located in the UKIDSS LAS, 
	is the most distant quasar currently known.
	\citet{Yang2019} discovered four \(z>6.5\) quasars, 
	including one object with \(z=7.02\),
	using photometric data from the Dark Energy Survey (DES),
	the VISTA Hemisphere Survey (VHS)
	and the Wide-field Infrared Survey Explorer (\textit{WISE}).
	\citet{Wang2018b} recently published the first broad-absorption line quasar at \(z>7\)
	using photometric data from the 
	Dark Energy Spectroscopic Instrument Legacy Survey (DELS),
	Pan-STARRS1 and \textit{WISE}.
	Finally, a faint (\(M_{1450} = -24.13\)) \(z>7\) quasar 
	has been discovered using data from the Subaru HSC \citep{Matsuoka2019}.
	Further $z>6.5$ quasars have been discovered using NIR data from
	the UKIDSS LAS \citep{Wang2017}, 
	the UKIDSS Hemisphere Survey \citep{Wang2018a},
	the VISTA Kilo-Degree Infrared Galaxy (VIKING) survey \citep{Venemans2013},
	Pan-STARRS \citep{Venemans2015a,Decarli2017,Koptelova2017,Tang2017},
	the VHS \citep[][]{Reed2017,Reed2019,Pons2019},
	and the Subaru HSC \citep{Matsuoka2016,Matsuoka2018a,Matsuoka2018b}.
	
	Quasars at \(z>7\) are particularly valuable for exploring the epoch of reionisation.
	Absorption in the \lya forest saturates at very low values of the volume averaged cosmic neutral hydrogen fraction,
	$\bar{x}_{\rm HI} > 10^{-4}$, and this technique ceases to be a useful probe
	of reionisation at redshifts much greater than six \citep{Barnett2017}. 
	Detection of the red damping wing of the IGM 
	can be used to measure the cosmic neutral fraction when the Universe is substantially neutral, 
	$\bar{x}_{\rm HI}>0.05$. 
	Detection of this feature has been reported for two \(z>7\) quasars,
	suggesting that the neutral fraction rises rapidly over the redshift interval $6<z<7$. 
	The first \lya damping wing measurement was made in the spectrum of the  \(z=7.08\) quasar ULAS J1120+0641, 
	by \citet{Mortlock2011}, who found a neutral fraction of $\bar{x}_{\rm HI} >0.1$. 
	This measurement was refined by \citet{Greig2017a,Greig2017b}, 
	who obtained  $\bar{x}_{\rm HI} = 0.40\substack{+0.21 \\ -0.19}$ (68 per cent range), 
	using an improved procedure for determining the intrinsic \lya emission-line profile.
	An even higher neutral fraction ($\bar{x}_{\rm HI} = 0.60\substack{+0.20 \\ -0.23}$)
	was obtained from analysis of the spectrum of the 
	\(z=7.54\) quasar ULAS J1342+0928, by \citet{Banados2018} and \citet{Davies2018}. 
	In contrast \citet{Greig2019} record a lower value $\bar{x}_{\rm HI} = 0.21\substack{+0.17 \\ -0.19}$ for this source. 
	Some uncertainty remains over the \lya damping wing measurements made to date, 
	given the difficulties associated with reconstructing the intrinsic \lya emission lines,
	and noting that these two $z>7$ quasars are not typical compared to lower-redshift counterparts,
	in that they both have large \Civ blueshifts.
	
	The picture of a substantially neutral cosmic hydrogen fraction at $7<z<8$ suggested by 
	these two $z>7$ quasars
	is in agreement with the latest constraints on reionisation 
	from measurements of the cosmic microwave background (CMB) by the \textit{Planck} satellite.
	Successive improvements of the measurement of 
	polarisation of the CMB have led to a progressive decrease in the best estimate of 
	the electron scattering optical depth, corresponding to an increasingly late EoR, 
	with the midpoint redshift of reionisation 
	most recently found to be \(z = 7.7\pm0.7\) \citep{Planck2018}.
	This motivates the discovery of a large sample of bright $z>7$ quasars, 
	and further development of methods for reconstructing the intrinsic \lya emission line, 
	to improve measurements of the \lya damping wing. 
	This will allow the progress of reionisation to be studied in detail. 
	
	Bright $z>7$ quasars will also be useful in other ways for studying the EoR.
	For example,
	assuming the measured decline in near zone sizes 
	with redshift continues \citep{Carilli2010,Eilers2017},
	the resulting \lya surface brightness 
	of the quasar Str\"{o}mgren sphere may be detectable \citep{Cantalupo2008,Davies2016},
	allowing detailed study of the structure of the mostly neutral IGM.
	Finally, extending measurements of the QLF beyond \(z=7\) 
	will be important for SMBH growth models,
	and will allow us to quantify the contribution of AGN
	to the earliest stages of reionisation,
	a topic of recent interest following the possible X-ray detection
	of faint active galactic nuclei candidates at \(z>4\) by \citet{Giallongo2015}
	\citep[but for a different interpretation of the same data, see][]{Parsa2018}.
	
	The prospects for finding many more bright 
	\(z>7\) quasars in the short term, using existing datasets, are poor nevertheless.
	The main reason for this is simply that \(z>7\) quasars are very rare:
	for example, assuming \(k=-0.72\) is applicable at \(z>6\),
	the results of \citet{Jiang2016} imply there are only \(\sim\!200\) redshift \(7<z<9\) quasars 
	brighter than \(J_{\rm AB} = 22\) over the whole sky. 
	In the redshift interval \(7<z<9\), \lya lies in the \(Y\) or \(J\) band. 
	To discriminate against contaminants requires one or more bands redward of the \lya band,
	so optical surveys, including those 
	stretching to the \(Y\) (or \(y\)) band, such as DES and (in the future) LSST, 
	are not competitive on their own. 
	Multiband, deep, widefield, near-infrared surveys, combined with deep optical data are ideal.
	
	Existing near-infrared datasets
	such as the LAS, VHS, and VIKING
	have been thoroughly searched,
	but do not survey a sufficient volume to yield significant numbers of bright sources.
	Selection of \(z>7\) quasars is hampered 
	by contamination from intervening populations:
	late M stars, and L and T dwarfs (hereafter MLTs); 
	and early-type galaxies at \(z=\myrange{1}{2}\), which we also refer to as `ellipticals' in this work. 
	These populations are far more common than,
	and have similar NIR colours to, the target quasars \citep[e.g.][]{Hewett2006}.
	Consequently, colour-selected samples of fainter candidates 
	become swamped by contaminating populations, 
	especially as quasar searches move to lower \sn to maximise the number of discoveries.
	
		\begin{figure}[t]
		\centering
		\includegraphics[width=6cm,trim={0 0.40cm 0 0.3cm}]{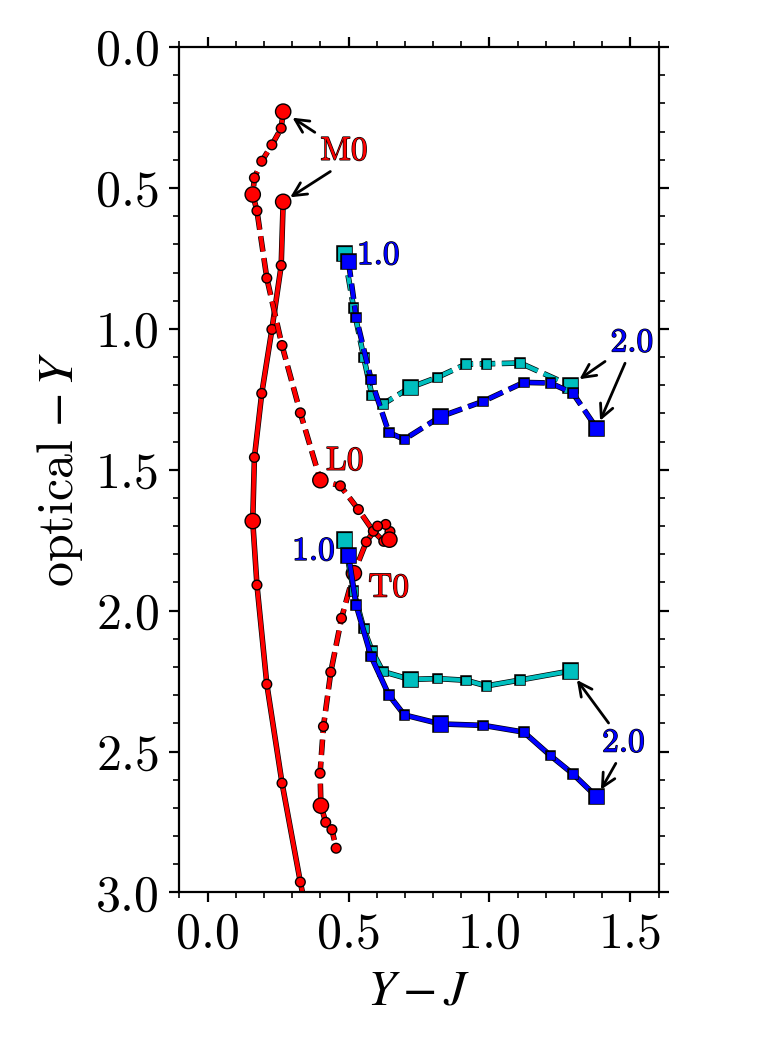}
		\includegraphics[width=6cm,trim={0 0.40cm 0 0}]{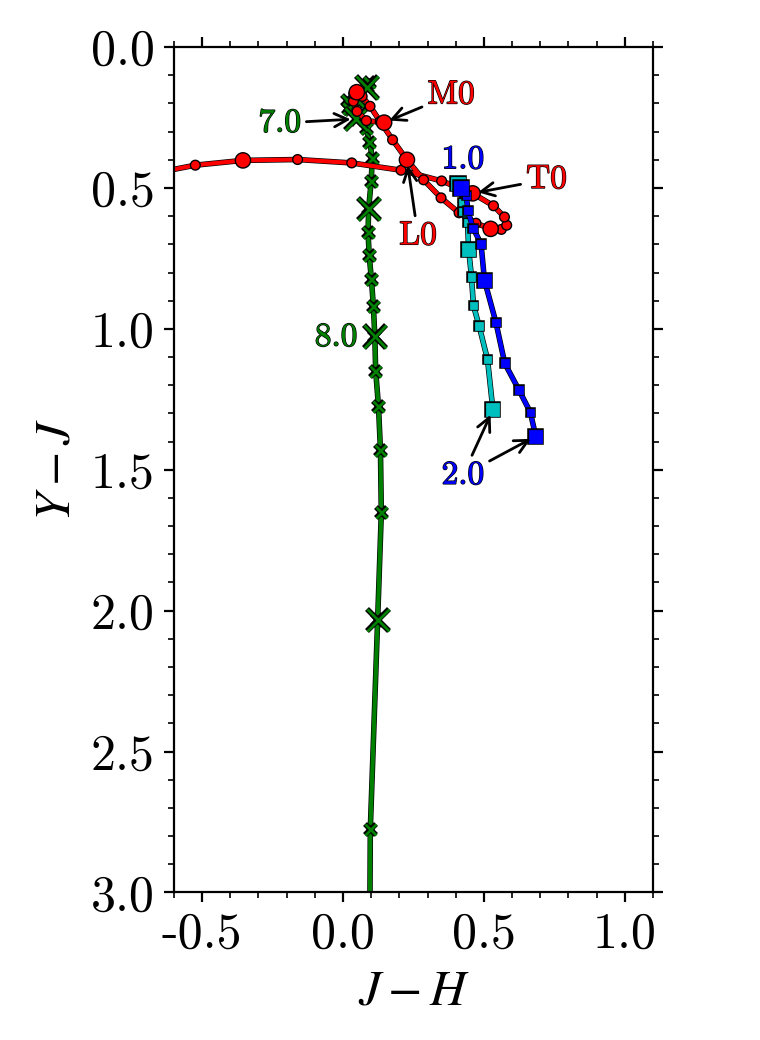}
	   \caption{Model colour tracks of relevant populations.
	   	We describe the population modelling in Sect.~\ref{sec:pops}.
	   	The separate populations are indicated as follows.
	   	The red tracks with circles show model MLT colours 
	   	for each spectral type.
	   	The blue tracks with squares indicate 
	   	early-type elliptical populations
	   	with two formation redshifts
	   	(\(\zf=3\), light blue;  and \(\zf=10\), dark blue),
	   	with spacing \(\Delta z = 0.1\), and redshift labels.
	   	The green track with crosses 
	   	indicates quasar model colours, 
	   	with spacing \(\Delta z = 0.1\), 
	   	and redshift labels.
	   \textit{Upper:} Optical-\(YJ\) colours.
	   	\(z>7\) quasars are expected to have negligible flux in both \(O\) and \(z\),
	   	so would appear below the bottom of the plot. 
	   		We present separate tracks for the two optical bands.
	    Solid lines indicate where the \euc \(O\) band is used in the optical.
	   Dashed lines indicate where the ground-based \(z\) band is used in the optical instead.
	   \textit{Lower:} \euc \(YJH\) colours.}
		\label{fig:colourcolour}
	\end{figure}	
	
	The launch of \euc, currently planned for Q2 2022, 
	should prove to be a landmark in high-redshift quasar studies.
	An analysis of potential quasar yields in the \euc wide survey
		was previously carried out for the Red Book \citep[][sect. 2.4.2]{Laureijs2011},
		based on cuts in \(YJH\) colour space.
	    That report focused especially on \(z>8.1\) quasars,
	    which are much redder than the contaminants in \(Y-J\), and so may be separated on that basis
	    (see \citealt{Laureijs2011}, Figure 2.6).
	    In contrast, over \(7.2 \lesssim z \lesssim 8.1\), 
	    near infrared broadband colours cannot easily separate quasars from contaminating populations, 
	    except with very deep complementary $z$-band data.
	    Since then, the \euc Near Infrared Spectrometer and Photometer (NISP) 
	    instrument wavelengths have changed \citep{Maciaszek2016}.
	    In particular, this has resulted in bluer \(Y-J\) colours for the three populations than was the case in \citet{Laureijs2011}.
	    We show the revised model colour tracks of the three populations that we consider in this work in Fig.~\ref{fig:colourcolour}.    
	    
	Contamination becomes more of a problem at low S/N,
	which was dealt with in the \citet{Laureijs2011} analysis 
	by selecting only bright point sources (\(J_\mathrm{AB}<22\)).
	Furthermore, it was argued that early-type galaxies at these brighter magnitudes might be identified 
	and eliminated on the basis of their morphologies (we examine this assumption in more detail below).
	Assuming the \(z=6\) QLF of \citet{Willott2010}, with \(k = -0.47\),
	it was predicted in \citet{Laureijs2011} that 30 \(z>8.1\), 
	\(J_\mathrm{AB}<22\) quasars would be found in the \(15\,000\,\dsq\) wide survey.
	
	Adopting the rate of decline \(k=-0.72\) measured by \citet{Jiang2016}, 
	has a dramatic effect on the predicted numbers in \citet{Laureijs2011}, 
	reducing the yield of \(z>8.1\) quasars from 30 to just eight. 
	As already noted, the real situation may be worse than this, 
	if the acceleration of the decline measured over $4<z<6$ continues beyond $z=6$. 
	But if finding high-redshift quasars in \euc will be more difficult than previously thought, 
	this is true for all surveys, 
	and the \euc wide survey remains by far the best prospect for searches for high-redshift quasars. 
	This motivates a deeper study of the problem, 
	and reconsideration of the prospects for finding quasars in \euc in the redshift interval $7<z<8$, 
	as well as for finding fainter quasars ($J_\mathrm{AB}>22$) at $z>8$. 
	The aim of this paper, therefore, is to improve on the \citet{Laureijs2011} analysis,
	and update predicted quasar numbers in the \euc wide survey over the entire redshift interval $7<z<9$.
	To this end, we have developed better models of the contaminating populations,
	and we also explore more powerful selection methods which allow us to go fainter.
	We also consider the impact of deep ground-based \(z\)-band optical data on the predicted numbers.
    
    The aim of this paper is to accurately model the high-redshift quasar selection process,
    and make robust predictions of the \euc quasar yield,
    appropriate for the \euc Reference Survey (ERS)
    currently defined in Scaramella et al. (in prep.).
    We compare selection using either \euc or \(z\)-band optical data,
    focusing in particular on the overwhelming contamination from MLTs and early-type galaxies.
	The paper is structured as follows. 
	We summarise the data that will be available to us,
	both from \euc and from complementary ground-based surveys, in Sect.~\ref{sec:surveys}. 
	We then describe the methods that we use to select \(z>7\) quasars,
	and the population models that underpin them in Sect.~\ref{sec:selection}.
	In Sect.~\ref{sec:qsf} we present the results of simulations of high-redshift quasars, 
	in the form of quasar selection functions, 
	i.e., detection probabilities as a function of absolute magnitude and redshift,
	and the corresponding predicted numbers of quasars that will be discovered.
	In Sect.~\ref{sec:discuss} we discuss the main uncertainties which will bear on the ability
	to select high-redshift quasars in the wide survey,
	and additionally discuss a potential timeline for \euc \(z>7\) quasar discoveries.
	We summarise in Sect.~\ref{sec:sum}. 
	We have adopted a flat cosmology 
	with \(h\) = 0.7, \(\Omega_{\rm m}\) = 0.3, and \(\Omega_{\Lambda}\) = 0.7.
	All magnitudes, colours and k-corrections quoted are on the AB system, 
	and we drop the subscript for the remainder of the paper. 
	
	\begin{table*}[t] 
	\centering
	\advance\leftskip-3cm
	\advance\rightskip-3cm
	\caption{Summary of survey combinations explored in simulations in this paper.}
	\label{tab:surveys}
	\begin{tabular}{ccccc}
		\hline \hline \\[-2ex]
		Survey(s)  & Depth in near-infrared & Depth in optical & Positional constraints & Fiducial area  \\ 
		\hline \\[-2ex]
		\euc       & \(YJH\)  24.0  ($5\,\sigma$) & \( O \) 24.5  ($10\,\sigma$)    & ERS coverage (Fig.~\ref{fig:sky}) & \(15\,000\,\dsq\) \\
		\hline\\[-2ex]
		\euc + PS (DR3) & \(YJH\) 24.0 ($5\,\sigma$) &   \(z\)  24.5 ($5\,\sigma$) &  as \euc only, and \(\delta > 30 \degree\) & \(5\,000\,\dsq\) \\
		\euc + LSST (1 yr) & \(YJH\) 24.0 ($5\,\sigma$) &   \(z\) 24.9 ($5\,\sigma$) &  as \euc only, and \(\delta < 30 \degree\) & \(10\,000\,\dsq\)	\\
		\hline		
	\end{tabular}
\end{table*}	
	\begin{figure}[t] 
		\centering
		\includegraphics[width=9cm]{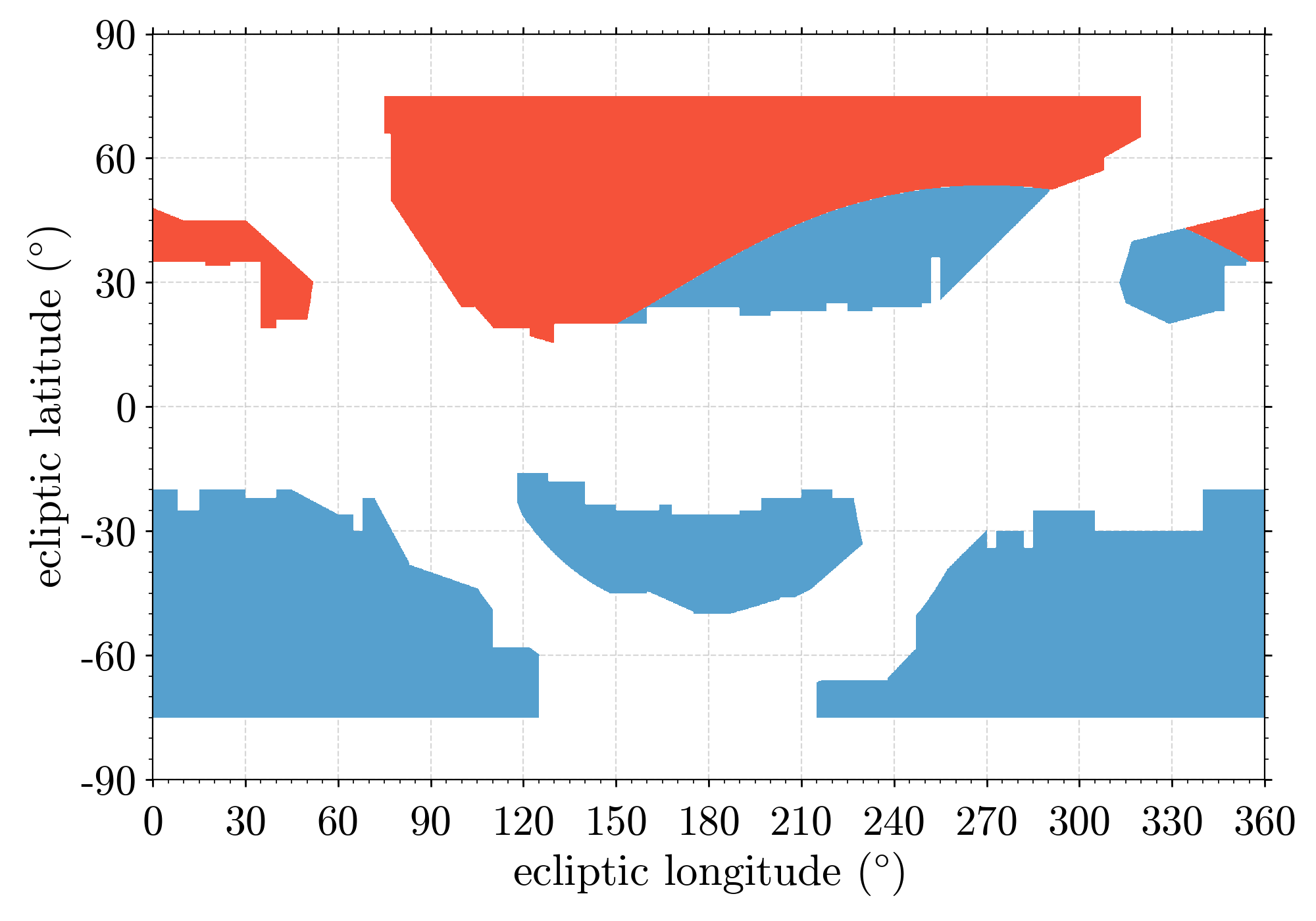}
		\caption{Cylindrical projection of the area from which we draw simulated quasars, in ecliptic coordinates,
			consistent with the ERS coverage defined in Scaramella et al. (in prep.).
			\euc/Pan-STARRS sources are drawn from the red area with \(\delta > 30 \degree\), 
			and \euc/LSST sources from the blue area with \(\delta < 30 \degree\). 
			The sample with no ground-based counterpart is drawn from the combined area.}
		\label{fig:sky}
	\end{figure}

	\section{Data}
	\label{sec:surveys}
	
	In Sect.~\ref{sec:euclid} we give a brief technical overview of the \euc wide survey
	(see \citealt{Laureijs2011} and Scaramella et al., in prep., for further details).
	The search for high-redshift quasars is enhanced with deep data in the $z$ band, 
	and we summarise complementary ground-based optical data in Sect.~\ref{sec:ground}. 
	The areas and depths of the \euc and ground-based data are summarised in Table~\ref{tab:surveys}. 
	
	Since the ground-based data have not yet been secured, in this paper two scenarios are considered:
	i) the case where \euc data are the only resource available,
	for which we consider optical data from the visual instrument, 
	in a wide filter (\(R+I+Z\)) which for brevity we label $O$;
	and ii) where we replace \euc optical data with complementary ground-based \(z\)-band data.

   	\subsection{\euc wide survey}
	\label{sec:euclid}
	
	The \euc wide survey will offer an unprecedented resource for \(z>7\) quasar searches,
	in terms of the combination of area covered and the NIR depths achieved. 
	The six-year wide survey of \euc will cover \(15\,000\,\dsq\) of extragalactic sky
	in four bands: a broad optical band (\(O\); \myrange{5500}{9000}\,\AA);
	and three near-infrared bands, \(Y\) (\myrange{9650}{11\,920}\,\AA), 
	\(J\) (\myrange{11\,920}{15\,440}\,\AA) and \(H\) (\myrange{15\,440}{20\,000}\,\AA).
	The planned depths, from \citet{Laureijs2011}, are provided in Table~\ref{tab:surveys}.
	
	The exact sky coverage of the \euc wide survey is yet to be finalised,
	with multiple possible solutions which satisfy the minimum area and science requirements laid out by \citet{Laureijs2011}.
	The assumed sky coverage is relevant to this paper, 
	because the surface density of MLT dwarfs depends on Galactic latitude (Sect.~\ref{sec:ms}).
	To ensure the results of this paper accurately reflect quasar selection with \euc,
	we follow the ERS shown in Scaramella et al. (in prep.),
	additionally indicated in Fig.~\ref{fig:sky}\footnote{
	In addition to the assumed ERS coverage,
		we simulate quasar selection in \euc assuming alternative wide survey footprints,
		which satisfy the \euc area and science requirements.
		We find selection is not sensitive to these different versions of the wide survey,
		with the resulting quasar number count predictions simply scaling with area.
		That is to say, for a fixed area, the results of this paper are highly robust
		to the wide survey coverage specifics
		and quasar selection will not change with future iterations of the ERS.
	}.
	We assign random sky coordinates drawn from the wide survey to all sources that we simulate.
	
	The fields are located at high Galactic latitudes, and so the reddening is low. 
		It is estimated that reddening $E(B-V)$ exceeds 0.1 over only $\myrange{7}{8}\%$ of the area \citep{Galametz2017}. 
		Any small regions of significantly higher reddening will be excised from the search for quasars. 
		For the remainder of the survey, with $E(B-V)<0.1$, the effect on the quasar search is very small. 
		At this level of reddening the change in $Y-J$ colour of a quasar is 0.05. 
		This degree of reddening is within the range of colour variation of normal quasars. 
		The discrimination against MLT dwarfs and early-type galaxies will not be affected 
		at this level of reddening since it is primarily set by the contrast at the Lyman break, which is barely changed.
		
		The \euc deep fields will ultimately reach two magnitudes fainter
			than the wide survey in \textit{OYJH}; 
			however, they are unlikely to prove such a useful resource for \(z>7\) quasar searches,
		    since they will cover a total area of just \(40\,\dsq\) (Scaramella et al. in prep.).
	        Without any consideration of the completeness of a potential search for quasars in the deep survey,
	        the \citet{Jiang2016} QLF (\(k=-0.72\)) implies
	        12 sources with \(7<z<9\) in the magnitude range \(J=\myrange{24}{26}\),
	       i.e., in the deep survey but not the wide survey. 
	       As shown in Sect.~\ref{sec:bmcresults}, this is a factor of \myrange{10}{20} lower 
	       than the predicted yield of \(J\lesssim23\) quasars from the wide survey,
	       depending on the choice of optical data used to select candidates.
	       Spectroscopic follow-up of \(J > 24\) quasar candidates 
	       in the deep fields will also prove challenging.
           Consequently, this work considers the \euc wide survey only.

	\subsection{Ground-based \(z\)-band optical data}
	\label{sec:ground}
	
	Sufficiently deep $z$-band data
	enhance the contrast provided across the \lya break in quasar spectra, 
	compared to the $O$ band (Fig.~\ref{fig:colourcolour}). 
	At redshifts $z>7$, there is negligible flux blueward of the \lya emission line in quasar spectra, 
	meaning quasars appear below the bottom of the upper panel of Fig.~\ref{fig:colourcolour}.
	The $z-Y$ colours of the potential contaminating populations, 
	MLTs, and early-type galaxies, are less red than for the quasars. 
	The large width of the $O$ band softens the contrast 
	between the colours of quasars and the colours of the contaminating populations.
	
	We wish to evaluate the extent to which using deep complementary \(z\)-band data
	from LSST \citep{Ivezic2008} and Pan-STARRS \citep{Chambers2016}
	can improve quasar selection over \(z=\myrange{7}{8}\).
	The current goal is that the entire wide survey area will be covered by a combination of 
	Pan-STARRS in the northernmost \(5000\,\dsq\) of \euc sky,
	with the remaining area covered by LSST \citep{Rhodes2017}.
	We therefore concentrate on these two ground-based resources.
	
	The exact crossover areas between \euc and the ground-based surveys, 
	and the target \(z\)-band depths are still to be finalised,
	so we have made a set of working assumptions for the purposes of this paper,
	which are summarised in Table~\ref{tab:surveys}.
	We additionally indicate the assumed crossover area in Fig.~\ref{fig:sky}.
	The adopted \(5\,\sigma\) depth for LSST, $z = 24.9$, 
	is based on one year of data 
	(following the start of operations scheduled for 2022),
	assuming 20 zenith observations of each source 
	with a single-visit \(5\,\sigma\) depth of \(z=23.3\) \citep{Ivezic2008}.
	The proposed LSST crossover area is composed of three separate surveys:
	the LSST main survey covering \(-62 \degree < \delta < 2 \degree\), 
	and northern (\(2 \degree < \delta < 30 \degree\)) 
	and southern (\(-90 \degree < \delta < -62 \degree\)) extensions,
	across which the final depths will differ \citep{Rhodes2017};
	however, for the sake of simplicity, we assume a uniform LSST depth.
	For Pan-STARRS we assume the planned depth at the time of \euc DR3 (2029),
	which is anticipated to be \(z=24.5\) at \({\rm S/N} = 5\).
	The Pan-STARRS and LSST \(z\) filter curves are extremely similar,
	and the resulting \(z-Y\) colours are essentially identical.
	Differences in the selection functions for the two surveys are driven by the different depths in \(z\).
		Although not considered further here,
		we note the possibility of using \(z\)-band data from other sources in the future, e.g., DES,
		which will cover \(5000\,\dsq\) of sky, almost entirely in the southern celestial hemisphere, in common with LSST.
		DES is ultimately expected to reach a \(5\,\sigma\) depth of \(z\sim24\) \citep[e.g.][]{Morganson2018},
		i.e., around 1\,mag shallower than LSST; 
		however, DES data will be available considerably sooner,
		with the final release (DR2) expected August 2020.
			
	
	\section{High-redshift quasar selection}
	\label{sec:selection}
	
	In this work, predictions of quasar numbers from the \euc wide survey
	are based on quasar selection functions,
	which reflect the sensitivity of \euc to quasars
	using a particular selection method
	as a function of luminosity and redshift,
	and over which different QLFs can be integrated to determine quasar yields.
	The starting point is a large number of simulated quasars 
	on a grid in luminosity/redshift space.
	We simulate realistic photometry for these sources
	using model colours (Sect.~\ref{sec:mq}),
	and add Gaussian noise to the resulting fluxes 
	based on the assumed depths in each band.
	We determine selection functions by recording the proportion recovered
	when given selection criteria are applied to the sample.
	For computing the selection function, 
	the details of the completeness of the \euc catalogues around the detection limit \(J\sim24\), 
	are unimportant because we find the efficiency of 
	the selection algorithm falls rapidly fainter than \(J\sim 23\) (Sect.~\ref{sec:qsf}).
	As such we do not simulate the full \euc detection process using the \(Y+J+H\) stack,
	and require only that a source be measured with \(J<24\) before we apply the selection criteria.
		
	The analysis provided in \citet{Laureijs2011} was based on colour cuts, indicated in Fig.~\ref{fig:colourcolour}. 
	This is an inefficient method as it does not weight the photometry in any way,
	and the chosen cuts are heuristic. 
	Here, instead, we employ and compare two different statistical methods for selecting the quasars. 
	These are described in Sect.~\ref{sec:methods}. 
	The first uses an update to the Bayesian model comparison (BMC) technique laid out by \citet{Mortlock2012}. 
	The second uses a simpler minimum-\(\chisq\) model fitting method (sometimes called `spectral energy distribution, or SED fitting'), 
	very similar to the method of \citet{Reed2017}.
	The methods are based on improved population models 
	for the key contaminants: MLT dwarf types; and compact early-type galaxies.
	Both methods require model colours for each population. The
	BMC method additionally requires a model for the surface density of each source as a function of apparent magnitude.
	We present the population models in Sect.~\ref{sec:pops}.
		In this work we assume that MLTs and early-type galaxies 
		are the only relevant contaminating populations
		for the selection of high-redshift quasars in \euc. 
		In Sect.~\ref{sec:cosmosdiscuss} we consider this assumption further,
		by analysing deep COSMOS data \citep{Laigle2016}.
		We do not see evidence for further populations that need be considered 
		for high-redshift quasar searches with \euc.
	
	\subsection{Selection methods}
	\label{sec:methods}
	
	We will now describe the two methods which we use to select candidate high-redshift quasars.
	The BMC method is presented in Sect.~\ref{sec:bmc},
	and the minimum-\(\chisq\) model fitting in Sect.~\ref{sec:chioverview}.
	Both methods are based on linear fluxes and uncertainties in each photometric band,
	even where a source is too faint to be detected.
	As such, we require some form of list-driven photometry,
	i.e., forced aperture photometry in all bands,
    for all sources that satisfy given initial criteria. 
	
	\subsubsection{Bayesian model comparison}
	\label{sec:bmc}
	
	The BMC method used in this work is principally the same as that proposed by \citet{Mortlock2012},
	which was used to discover the \(z=7.08\) quasar \uj.
	The crux of the method is to calculate a posterior quasar probability, \pq,
	for each source in a given sample,
	which allows candidates to be selected and prioritised for follow-up. 
	In short, \pq is given by the ratio of `weights' (\(W_t\)) of each type
	of object \(t\) under consideration.
	\citet{Mortlock2012} presented a general form for the calculation of \pq
	given any number of relevant populations,
	which in this case we take to be
	quasars, denoted \(q\); MLTs, \(s\); and early type galaxies, \(g\). 
	Explicitly, given a set of photometric data \({\bm d}\),
	\begin{equation}
	\label{eq:combw}
	P_{\mathrm q} \equiv p\left(q\given \bm{d}\right) = 
	\frac{W_\mathrm{q}\left(\bm{d}\right)}
	{W_\mathrm{q}\left(\bm{d}\right) + W_\mathrm{s}\left(\bm{d}\right) + W_\mathrm{g}\left(\bm{d}\right)}.
	\end{equation}
	To calculate the individual weights for a population, $q$, $s$, or $g$,
	we simultaneously make use of all available photometric data for a source,
	combined with the surface density of the population,
	which serves as a prior. 
	For any source, 
	the weight of a particular population measures the relative probability 
	that the source would have the particular measured fluxes, 
	assuming it to be a member of that population, 
	characterised by the model colours, 
	and surface density as a function of apparent magnitude. 
	\citet{Mortlock2012} applied the method to the case of two populations: quasars and M stars. 
	Here we adopt the more general form of the method, and apply it to three populations,
	which we describe fully in Sect.~\ref{sec:pops}. 
	The MLT population itself is divided into a set of sub-populations, 
	which are the individual spectral types from M0 to T8.
	This approach to the cool dwarf population is similar to that of \citet{Pipien2018},
	who developed models for each spectral type \myrange{L0}{T9}
	in a search for high-redshift quasars 
	in the Canada-France High-\(z\) Quasar Survey in the Near-Infrared.
	
	The individual weights for each population are calculated as follows:
	\begin{equation}
	\label{eq:gnrlweight}
	W_t\left(\bm{d}\right) =
	\int \Sigma_t\left(\bm{\theta}_t\right)\;
	p\left(\bm{d}\given \bm{\theta}_t,t\right)\;
	\diff\bm{\theta}_t,
	\end{equation}
	where \(\bm{\theta}_t\) is the set of parameters describing a single population. 
	The two terms in the integral in Eq.\,(\ref{eq:gnrlweight}) are respectively 
	the surface density function, 
	and a Gaussian likelihood function based on model colours,
	which is written in terms of linear fluxes.
	Explicitly, the full likelihood function is given by
	\begin{equation}
	\label{eq:lik}
    p\left(\bm{d}\given \bm{\theta_{t}}, t\right) = 
    \prod_{b=1}^{N_\mathrm{b}}\,
    \frac{1}{\sqrt{2\pi}\,\hat{\sigma}_b}\,
    \exp\left\{-\frac{1}{2} 
    \left[\frac{\hat{f}_b - f_b\left(\bm{\theta_{\emph t}}\right)}
    {\hat{\sigma}_{\emph b}}\right]^2\right\},
	\end{equation}
	where the data in each of the \(N_\mathrm{b}\) bands \(b\) is of the form \(\hat{f}_b\pm\hat{\sigma}_b\), 
	and \(f_{b}\left(\bm{\theta_{\emph t}}\right)\) is the true flux in band \(b\) of an object of type \(t\) 
	described by the parameters \(\bm{\theta_t}\).
        From the above definition of the individual population weights, 
        which incorporates both the prior weighting and likelihood, 
        it follows that the ratio of any pair of population probabilities (\(P_{\mathrm q}, P_{\mathrm g}, P_{\mathrm s}\); cf. Eq.~\ref{eq:combw}) 
        yields the product of a prior ratio and a Bayes factor \citep[e.g.][]{Sivia2006}.

	The chosen threshold value of $P_{\mathrm q}$ that defines the sample of candidate quasars, 
	effects a balance between contamination and completeness. 
	In this work the selection functions are computed for a probability threshold of \(P_{\mathrm q}=0.1\), 
	consistent with \citet{Mortlock2012}. 
	This implies a follow-up campaign to identify unambiguously all sources with \(P_{\mathrm q}>0.1\), e.g., with spectroscopy. 
	The value \(P_{\mathrm q}=0.1\) was chosen initially 
	because it worked well for the UKIDSS LAS high-redshift quasar survey \citep{Mortlock2012}\footnote{
    In the \citet{Mortlock2012} survey, \(P_{\mathrm q}=0.1\) was chosen as the selection criterion for visual inspection,
	which resulted in a candidate list of 107 real objects.
	Of these, the discovered quasars typically had much higher probabilities:
	in total there were 12 \(z\gtrsim6\) quasars discovered in UKIDSS (or previously known from SDSS), 
	of which ten had \(P_{\mathrm q}>0.9\) and two had \(0.4<P_{\mathrm q}<0.5\) (see \citealt{Mortlock2012}, Figures 10 and 13).
    }.
	As a check we also carried out detailed simulations of the contaminating populations,
	i.e., we created a synthetic \euc survey, and classified all sources. 
	A small fraction of non-quasars are selected as quasars; 
	however, \(P_{\mathrm q}>0.1\) is sufficient to exclude the majority of contaminants. 
	We present a full discussion of the \euc contaminants in Sect.~\ref{sec:contam}.
	
	In practice, 
	the \pq threshold will be set to control the number of candidates which are accepted for follow-up observation, 
	based on the expected numbers of quasars, 
	and will depend among other things on the reliability of the \euc photometry, 
	and the extent to which non-Gaussian errors (from whatever cause) afflict the data. 
	A lower value of $P_{\mathrm q}$ can increase the quasar yield,
	at a cost of allowing greater contamination of the sample. 
	In fact we find the selection functions, and therefore the predicted yield,
	are not particularly sensitive to the choice of threshold
	as typically quasars are recovered with high probability.
	Therefore, foreseeably, any \pq threshold in the range \(\myrange{5}{20}\%\)
	could be chosen, 
		depending on the length of the actual candidate lists
	and the follow-up resources that are available\footnote{
		Preliminary investigations suggest adopting a threshold of \(P_{\mathrm q} = 0.05~(0.2)\)
			results in a \(\sim15\%\) increase (decrease) in the total quasar yield
			compared to the results presented in Sect.~\ref{sec:bmcresults},
			driven by changes in the numbers of \(z<8\) quasars near the survey detection limit.}.
	
	\subsubsection{\(\chisq\) model fitting}
	\label{sec:chioverview}
	
	To assess the performance of the BMC method,
	we will also consider \euc quasar selection using a minimum-\(\chisq\) technique.
	Such an approach has previously been used by, e.g., \citet{Reed2017},
	who discovered eight bright (\(z_\mathrm{AB} < 21.0\)) $z\sim6$ quasars, 
	using a combination of DES, VHS and \textit{WISE} data.
	We calculate \(\chisq_{\rm red}\) values for a given source and model SED \(m\) as follows:
	\begin{equation}
	\chisq_{{\rm red,}m} = \frac{1}{N_\mathrm{b} - 2} 
	\sum_{b}^{N_\mathrm{b}}
	\left(\frac{\hat{f}_b - s_{\rm best}\,f_{m,b}}{\hat{\sigma}_b}\right)^2,
	\label{eq:chisq}
	\end{equation}
	where \(f_{m,b}\) is the (unnormalised) model SED flux in band \(b\),
	and \(s_{\rm best}\) is the normalisation that minimises \(\chisq\).
	We have \(N_\mathrm{b} - 2\) degrees of freedom as there are two parameters under consideration: 
	the normalisation of a single model and the range of models being fitted \citep[e.g.][]{Skrzypek2015}.
	(That is to say, for the quasars and early-type galaxies, the second parameter is redshift, while for the MLT dwarfs the second parameter is spectral type, since they form a continuous sequence.)
	The SED fitting can be linked to the BMC method by considering the logarithm of the likelihood given in Eq.\,(\ref{eq:lik}). 
	The key difference in the SED fitting method compared to the BMC method is that
	no surface density information is employed, i.e., we do not include a prior.
	
	We use the model colours outlined in Sect.~\ref{sec:pops} to produce quasar and contaminant SEDs,
	and fit them to the fluxes of each source, following Eq.\,(\ref{eq:chisq}).
	Therefore for the MLTs each spectral type represents a model,
	while for the galaxies and quasars the set of models is defined by SEDs produced in intervals of \(\Delta z = 0.05\).
	We keep the single best fitting quasar $(q)$ model and contaminant $(c)$ model, 
	with respective \(\chisq_{\rm red}\) values \(\chisq_{{\rm red,q(best)}}\)
	and \(\chisq_{{\rm red,c(best)}}\). 
	Following \citet{Reed2017}, 
	we apply two cuts to the \(\chisq_{\rm red}\) values to retain a source (see Figure 15 of that work).
	We firstly require \(\chisq_{{\rm red,c(best)}} > 10\), 
	i.e., the data are a bad fit to all contaminant models. 
	We additionally require the ratio \(\chisq_{{\rm red,c(best)}}/\chisq_{{\rm red,q(best)}} > 3\),
	i.e., the data are fit substantially better by a quasar SED than any contaminant model.
    In a similar way to the $P_{\mathrm q}$ threshold discussed in Sect.~\ref{sec:bmc},
    these cuts do not have a particular statistical significance,
    but would be chosen to control future candidate lists.
    It is likely that the optimal thresholds for \euc will ultimately differ from the \citet{Reed2017} study,
    due to differences in the data and the number of bands available, \(N_\mathrm{b}\).

	\subsection{Population models}
	\label{sec:pops}
	
	\begin{figure}[t] 
		\centering
		\includegraphics[width=7cm]{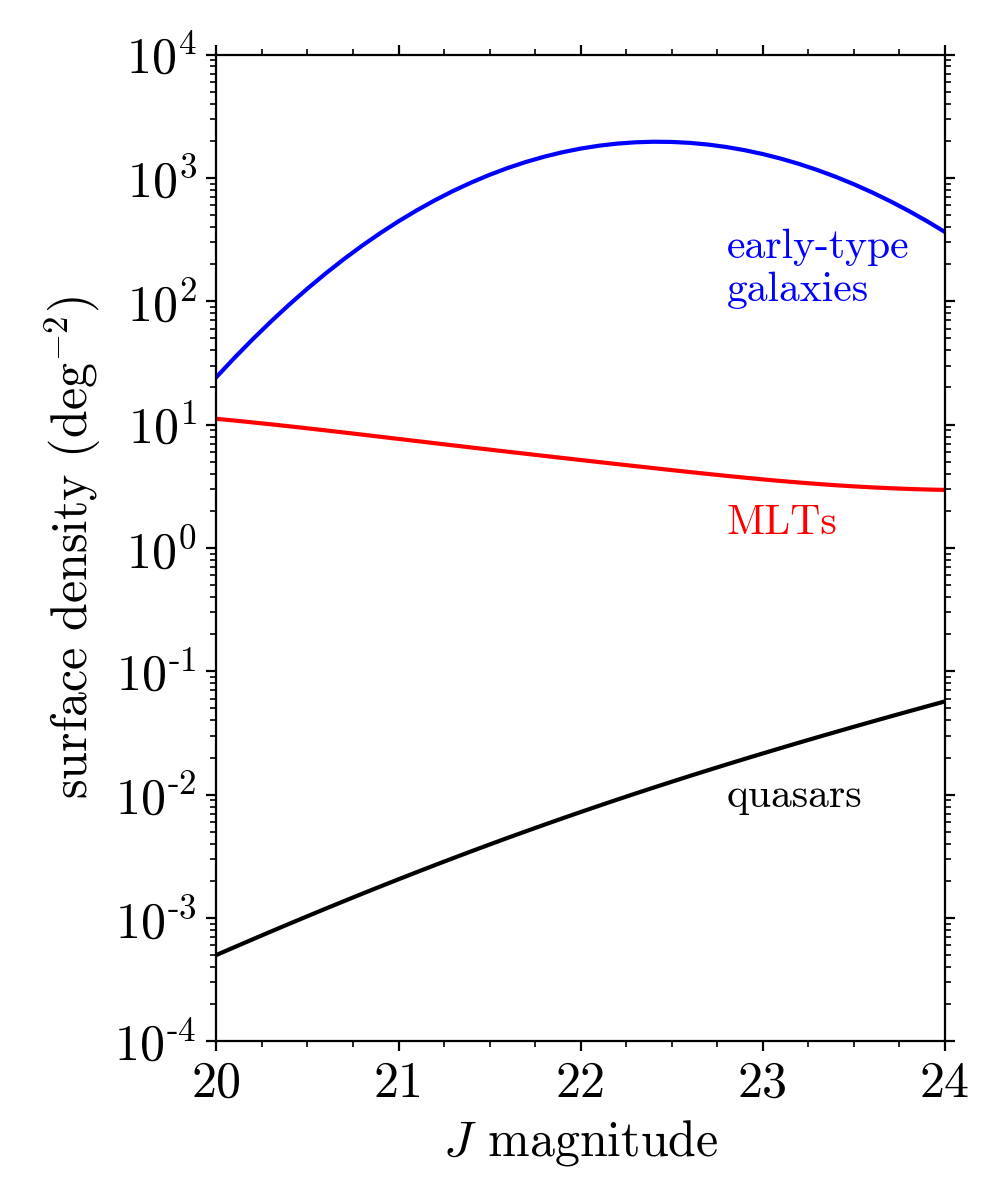}
		\caption{Population surface densities as a function of \(J\)-band magnitude.
		Blue: Early-type galaxies, integrated over the redshift range \(z=\myrange{1}{2}\).
		Red: MLTs, summed over spectral type.
		Black: Quasars, integrated over over the redshift range \(z=\myrange{7}{9}\).	
		}
		\label{fig:surfacedensity}
	\end{figure}
	
	We now summarise the surface density terms (Fig.~\ref{fig:surfacedensity}) and model colours (shown in Fig.~\ref{fig:colourcolour})
	which are used in the methods described in Sect.~\ref{sec:methods}.
	We present the models for quasars in Sect.~\ref{sec:mq},
	for MLTs in Sect.~\ref{sec:ms}, and for early-type galaxies in Sect.~\ref{sec:mg}.
	
	\subsubsection{Quasars}
	\label{sec:mq} 
	The parameters \(\bm{\theta}\) for the quasar weight $W_\mathrm{q}$ are absolute magnitude and redshift. 
	The surface density term is based on the \citet{Jiang2016} QLF,
	extrapolated to redshifts \(z>6\) assuming $k=-0.72$.
	We show the surface density of quasars in the redshift range \(7.0 \le z \le 9.0\)
		as a function of \(J\) magnitude in Fig.~\ref{fig:surfacedensity}.
		This helps to illustrate the difficulty associated with identifying quasars at \(z>7\),
	    as the quasar surface density is typically several orders of magnitude lower 
	    than those of both main contaminants.
	
	Quasar k-corrections and colours,
	which are required for both selection methods,
	are measured directly from model spectra using \texttt{synphot} \citep{Laidler2008}.
    The model SEDs were determined using quasars
    with redshifts $0.1\le z \le 4.0$ from the SDSS DR7 \citep{Schneider2010}.
    Optical 
    photometry from the SDSS was combined
    with near-infrared 
    data from UKIDSS \citep{Lawrence2007}
    and $W1$ and $W2$ photometry from \textit{WISE} \citep{Wright2010}
    to provide an extensive multi-wavelength data set. 
    The quasar SED model was then generated via a minimisation procedure 
    to determine parameter values for components including continuum power-law slopes,
    and emission-line equivalent widths.
    The model SED has seen extensive use \citep[e.g.][]{Hewett2006, Maddox2008, Maddox2012},
    and reproduces the median photometric properties of
    the SDSS DR7 quasars to better than 5\,per cent over the full redshift range $0.1\le z \le 4.0$.
     
   In this paper we use a single reference model to represent typical quasars. 
    This model is characterised by a set of emission line equivalent-width ratios.
    The standard line strength has rest frame equivalent width
    \(\mathrm{EW_{\Civ} = 39.1\,\AA}\) and UV continuum slope, 
    defined by the ratio of $f_\lambda$ at rest frame \(1315\,\AA\) and \(2225\,\AA\),
    \(f_{1315}/f_{2225}=1.0\). 
    Since we are only interested in redshifts $z>7$, 
    we assume that all flux blueward of \lya is absorbed for all sources that we simulate,
	except that we include a near zone of size 3\,Mpc (proper). 
	The results are insensitive to the choice of near-zone size.
%
	
    In the actual search of the \euc data we will adopt a set of quasar spectral types,
    characterised by variations in the continuum (redder/bluer), 
    and variations in the equivalent width of the reference \Civ emission line, 
    keeping all emission line ratios fixed.
	The surface density term (i.e., the prior) will be divided in proportion to the expected numbers, 
	based on our knowledge of quasars at lower redshifts. 
	The total quasar weight $W_\mathrm{q}$ is the sum of weights over the different types. 
	This inference strategy is essential to maximise the yield from \euc. 
	The goal of the current paper, to compute the expected yield of high-redshift quasars, is different, 
	and we can adopt a simpler strategy, and compute $W_\mathrm{q}$, and so $P_\mathrm{q}$, 
	by adopting a single typical spectral type. 
	Performing similar calculations for other surveys,
    the estimated yields are very similar for two scenarios: 
    firstly, using a single spectral model for the simulated population of quasars, 
    and the same model (i.e., colour track) for the selection algorithm; 
    secondly, using a range of quasar models, suitably weighted, for the simulated quasars, 
    and using the same range of models, and weights, in the selection algorithm. 
    This statement is only true if the single model adopted is typical,
    i.e., of average line width and continuum slope. 
    The reason for this is that objects with, 
    e.g., stronger (weaker) lines have a higher (lower) probability of selection,
    compared to the typical spectrum, and the corresponding gains and deficits approximately cancel. 
    Therefore a selection function weighted over the different spectral types 
    is very similar to the selection function computed for the average type.
    In addition, the computation of selection functions is considerably faster when a single SED is used,
    since one value of $W_\mathrm{q}$ is calculated for a single grid of simulated quasars (around 500,000 objects).
    Extending the problem to $n$ SEDs requires the calculation of $n$ individual quasar weights,
    for $n$	grids of quasars of different types.
    We consider this matter further in Sect.~\ref{sec:lx_cx} 
    where we investigate templates with different continuum slopes and line strengths.
    The analysis therein reinforces the above conclusion.

    Irrespective of the intrinsic quasar SEDs adopted, 
	neutral hydrogen along the line of sight will mean they have no significant flux 
	in bands blueward of the redshifted \lya line, 
	and so all standard search methods exploit the fact that they will be optical drop-out sources.
	This approach, however, ignores the possibility of gravitational lensing by an intervening galaxy 
	that both magnifies the quasar image(s) and directly contributes optical flux.  
	There have been theoretical predictions that the fraction of multiply imaged quasars in a flux-limited sample 
	could be up to ~30\% \citep{Wyithe2002}, 
	although empirically this fraction is closer to 1\% \citep{Fan2019}.
	It has been argued \citep{Fan2019,Pacucci2019} that this discrepancy 
	is because the optical flux from the deflector galaxies mean that lensed high-redshift sources are not optical drop-outs.  
	If this is the dominant effect then there would be an additional population of \(z > 7\) quasars beyond that considered here.  
	However, whether they would be detectable 
	depends on the numbers of contaminating sources with comparable optical-NIR colours, 
	which we do not explore in this work.

	\begin{figure}[t] 
		\centering
		\includegraphics[width=8.5cm]{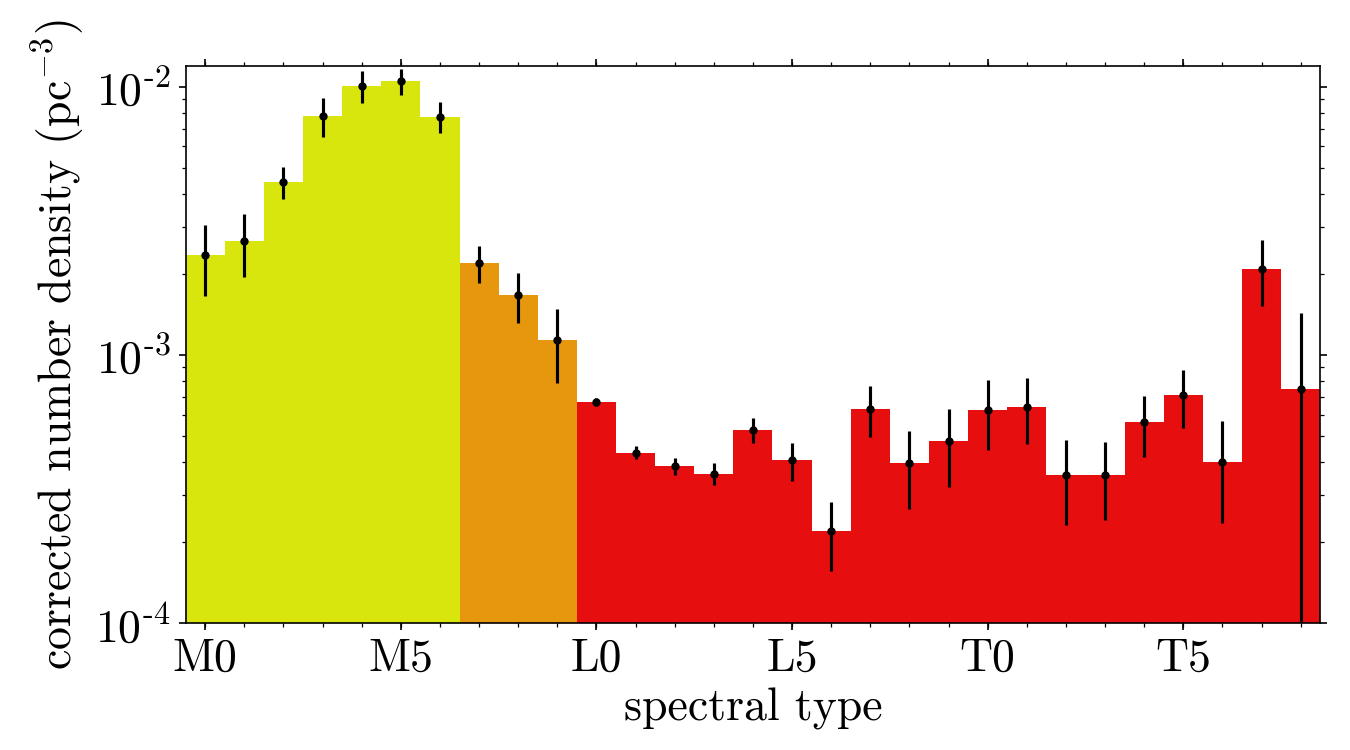}
		\caption{MLT number densities at the Galactic central plane.
			\myrange{M0}{M6} (yellow) are determined from the \citet{Bochanski2010} luminosity function.
			\myrange{M7}{M9} (orange) are extrapolated from L0, satisfying the \citet{Cruz2007} measurement.
			We measure \myrange{L0}{T8} (red) number densities from the \citet{Skrzypek2016} LAS sample.
		}
		\label{fig:mlt}
	\end{figure}
	
	\begin{table}[t] 
		\centering
		\advance\leftskip-3cm
		\advance\rightskip-3cm
		\caption{MLT density at the Galactic plane, and model colour data.
			\(z-Y\) is applicable to both LSST and Pan-STARRS.
		    We additionally show the MLT colours in Fig.~\ref{fig:colourcolour}.}
		\begin{adjustbox}{width=0.48\textwidth,totalheight=\textheight,keepaspectratio}
			\label{tab:mlt}
			\begin{tabular}{ccccccS[table-format=2.3]}
				\hline \hline \\[-2ex]
				SpT & \(\rho_0~(\mathrm{pc^{-3}})\) & \(M_J\) & \(z-Y\) & \(O-Y\) & \(Y-J\) & \(~J-H\)\\ 
				\hline \\[-2ex]
				M0  &  \(2.4\times10^{-3}\)  &  6.49  &  0.23  &  0.55  &  0.27  &  0.15  \\
				M1  &  \(2.7\times10^{-3}\)  &  7.07  &  0.29  &  0.78  &  0.26  &  0.08  \\
				M2  &  \(4.4\times10^{-3}\)  &  7.71  &  0.35  &  1.00  &  0.23  &  0.05  \\
				M3  &  \(7.8\times10^{-3}\)  &  8.28  &  0.41  &  1.23  &  0.19  &  0.04  \\
				M4  &  \(1.0\times10^{-2}\)  &  8.90  &  0.46  &  1.46  &  0.17  &  0.04  \\
				M5  &  \(1.1\times10^{-2}\)  &  9.53  &  0.52  &  1.68  &  0.16  &  0.05  \\
				M6  &  \(7.8\times10^{-3}\)  &  10.85  &  0.58  &  1.91  &  0.18  &  0.07  \\
				M7  &  \(2.2\times10^{-3}\)  &  11.66  &  0.82  &  2.26  &  0.21  &  0.10  \\
				M8  &  \(1.7\times10^{-3}\)  &  12.08  &  1.06  &  2.61  &  0.26  &  0.13  \\
				M9  &  \(1.1\times10^{-3}\)  &  12.33  &  1.30  &  2.96  &  0.33  &  0.18  \\
				L0  &  \(6.7\times10^{-4}\)  &  12.54  &  1.54  &  3.32  &  0.40  &  0.23  \\
				L1  &  \(4.3\times10^{-4}\)  &  12.79  &  1.56  &  3.32  &  0.47  &  0.29  \\
				L2  &  \(3.8\times10^{-4}\)  &  13.11  &  1.64  &  3.42  &  0.54  &  0.35  \\
				L3  &  \(3.6\times10^{-4}\)  &  13.50  &  1.72  &  3.51  &  0.59  &  0.41  \\
				L4  &  \(5.3\times10^{-4}\)  &  13.93  &  1.75  &  3.56  &  0.63  &  0.47  \\
				L5  &  \(4.1\times10^{-4}\)  &  14.38  &  1.75  &  3.55  &  0.65  &  0.52  \\
				L6  &  \(2.2\times10^{-4}\)  &  14.80  &  1.72  &  3.53  &  0.65  &  0.56  \\
				L7  &  \(6.3\times10^{-4}\)  &  15.17  &  1.69  &  3.51  &  0.63  &  0.58  \\
				L8  &  \(3.9\times10^{-4}\)  &  15.44  &  1.70  &  3.54  &  0.60  &  0.57  \\
				L9  &  \(4.8\times10^{-4}\)  &  15.63  &  1.76  &  3.63  &  0.56  &  0.53  \\
				T0  &  \(6.3\times10^{-4}\)  &  15.72  &  1.87  &  3.79  &  0.52  &  0.46  \\
				T1  &  \(6.4\times10^{-4}\)  &  15.74  &  2.03  &  4.00  &  0.48  &  0.35  \\
				T2  &  \(3.6\times10^{-4}\)  &  15.71  &  2.22  &  4.24  &  0.44  &  0.21  \\
				T3  &  \(3.6\times10^{-4}\)  &  15.69  &  2.41  &  4.47  &  0.41  &  0.03  \\
				T4  &  \(5.6\times10^{-4}\)  &  15.74  &  2.58  &  4.64  &  0.40  &  \(-0.16\)  \\
				T5  &  \(7.1\times10^{-4}\)  &  15.93  &  2.69  &  4.73  &  0.40  &  \(-0.36\)  \\
				T6  &  \(4.0\times10^{-4}\)  &  16.32  &  2.75  &  4.73  &  0.42  &  \(-0.52\)  \\
				T7  &  \(2.1\times10^{-3}\)  &  16.98  &  2.78  &  4.72  &  0.44  &  \(-0.64\)  \\
				T8  &  \(7.5\times10^{-4}\)  &  17.95  &  2.84  &  4.80  &  0.46  &  \(-0.65\)  \\
				\hline
			\end{tabular}
		\end{adjustbox}
	\end{table}

	\subsubsection{MLT dwarfs}
	\label{sec:ms}
	Most MLT dwarfs detected in the \euc
		 wide survey will be members of the Galactic thin disk. 
		 At the end of this section we also consider the possibility 
		 that members of the thick disk (larger scale height, lower metallicity) 
		 need to be considered as potential contaminants. 
	The number density of the thin disk population is assumed to vary as $\rho=\rho_0\,e^{-Z/Z_\mathrm{s}}$, 
	where $\rho_0$ is the number density of any spectral type \myrange{M0}{T8} at the Galactic central plane,
	$Z$ is the vertical distance from the plane, and $Z_\mathrm{s}$ is the scale height, 
	assumed to be 300\,pc \citep[e.g.][]{Gilmore1983}. 
	The small offset of the Sun from the Galactic central plane is ignored. 
	Each spectral type, or sub-population, is then specified by 
	the value of $\rho_0$, the absolute magnitude in the $J$ band, and the $zOYJH$ colours. 
	The values adopted are provided in Table~\ref{tab:mlt}. 
	In determining $W_\mathrm{s}$, weights are computed for each spectral type, 
	with the total weight \(W_\mathrm{s}\) given by a sum over types. 
	In this work, random coordinates are drawn from the \euc wide survey 
		(Sect.~\ref{sec:euclid}) for each simulated source,
	allowing us to fully incorporate Galactic latitude in the calculation of \(W_\mathrm{s}\).
	In the case of simulated MLTs (Sect.~\ref{sec:contam}),
	the coordinates that are drawn additionally preserve the dependence on Galactic latitude.
	
	We assigned colours for each spectral type by measuring colours 
	for suitable sources in the SpeX Prism Library \citep{Burgasser2014}, and selecting the median value for each spectral type.
	\citet{Holwerda2018} recently presented \euc NIR colours for the MLT population. 
	They took a different approach, by measuring colours for the  standard stars in the library; 
	however, these individual spectra do not extend sufficiently bluewards to measure optical colours. 
	In addition to the colours presented in Table~\ref{tab:mlt},
	we determine median SDSS \(riz\) colours for types \myrange{M0}{M6}, 
	using bright sources from the \citet{West2011} sample.
	These colours are required to compute number densities and absolute magnitudes as detailed below.
	
	Number densities for types \myrange{M0}{M6} are based on the luminosity function of \citet{Bochanski2010} as follows.
	Interpolating the model \(i-z\) colours, 
	we approximate a range in \(i-z\) for each spectral type
	using the range \(\brackets{\mathrm{SpT}-0.5,\mathrm{SpT}+0.5}\).
	The \(i-z\) colour evolves linearly over the early M types,
	and we simply extrapolate to K9 to determine the \(i-z\) range for M0.
	The M7 \(i-z\) colour, needed to define the M6 range, comes from \citet{Skrzypek2016}.
	Using the relation in \citet{Bochanski2010}, we convert the \(i-z\) range 
	for each spectral type into a range in \(M_r\).
	The last step is to interpolate the binned system luminosity function in \citet{Bochanski2010}.
	Integrating over the \(M_r\) range for each spectral type, 
	we finally obtain number densities in \({\rm pc^{-3}}\).
	
	The L- and T-type number densities were calculated using 
	the UKIDSS LAS LT sample presented by \citet{Skrzypek2016}. 
	For a particular spectral type, we computed the value of $\rho_0$ that reproduces the number of sources in the sample, 
	given the assumed scale height, the magnitude range of the sample, 
	and the solid angle of the survey as a function of Galactic latitude.
	For \myrange{M7}{M9} we use \citet{Cruz2007}, 
	who measure a total number density of 
	\(4.9 \times 10^{-3}\,{\rm pc^{-3}}\) 
	for these three spectral types. 
	We approximate the individual number densities 
	by assuming the number density varies linearly across the range \myrange{M7}{L0},
	constrained such that the total \citet{Cruz2007} number density is reproduced. 
	The number density as a function of spectral type is ploted in Fig.~\ref{fig:mlt},
	while the surface density summed over spectral types is shown as a function of \(J\) magnitude in Fig.~\ref{fig:surfacedensity}.
	\citet{Dupuy2012} provide \(J\)-band absolute magnitudes for spectral types \myrange{M6}{T8}. 
	For \myrange{M0}{M5} we use \citet{Bochanski2010} to determine \(r\)-band absolute magnitudes,
	again based on the \(i-z\) colour for each spectral type, 
	and use model colours to convert \(M_r\) to \(M_J\).
	
        \euc is sufficiently deep that we expect 
		the metal-poor MLT population of the thick disk, 
		i.e., ultracool subdwarfs \citep{Zhang2018}, to become important at faint magnitudes.
		Assuming that the thick disk population contributes 10\% of the stellar number density 
		at the Milky Way central plane, and has a scale height of 700\,pc \citep{Ferguson17},
		the expected number densities of thin and thick disk stars
		become comparable at vertical distances \(\sim 1200\)\,pc
		from the Galactic plane, meaning that the thick disk dwarfs need to be considered in addition to the thin disk dwarfs. 
		The luminosities in Table~\ref{tab:mlt} imply spectral types
		down to L3 will be observable with \(J < 24\) at such distances, 
		and so may become a comparable source of contamination at faint \euc magnitudes.
		However, in Sect.~\ref{sec:contam} we find the most important spectral types
		in terms of contamination are in the range \myrange{T2}{T4}.
		These types are only observable to distances of \(\sim 450\)\,pc with \(J<24\),
		meaning the equivalent subdwarfs are unlikely to be a significant source of contamination.		
		Additionally, the subdwarf population is bluer than MLTs in the thin disk,
		which will help with discrimination from quasars.
		This was determined using the L subdwarfs presented by \citet{Zhang2018}.
		For objects in Table 1 of that work that matched to the UKIDSS LAS,
		we measured the UKIDSS \(Y-J\) colours
		and compared against the template colours from \citet{Skrzypek2015} for each spectral type.
		We found L type subdwarfs are on average 0.24\,mag 
		bluer in \(Y-J\) than the corresponding L dwarfs, meaning that contamination by subdwarfs is much less of a concern.
		For these reasons we do not include the thick disk in our modelling of MLTs.

\begin{figure}[t] 
	\centering
	\includegraphics[width=8cm]{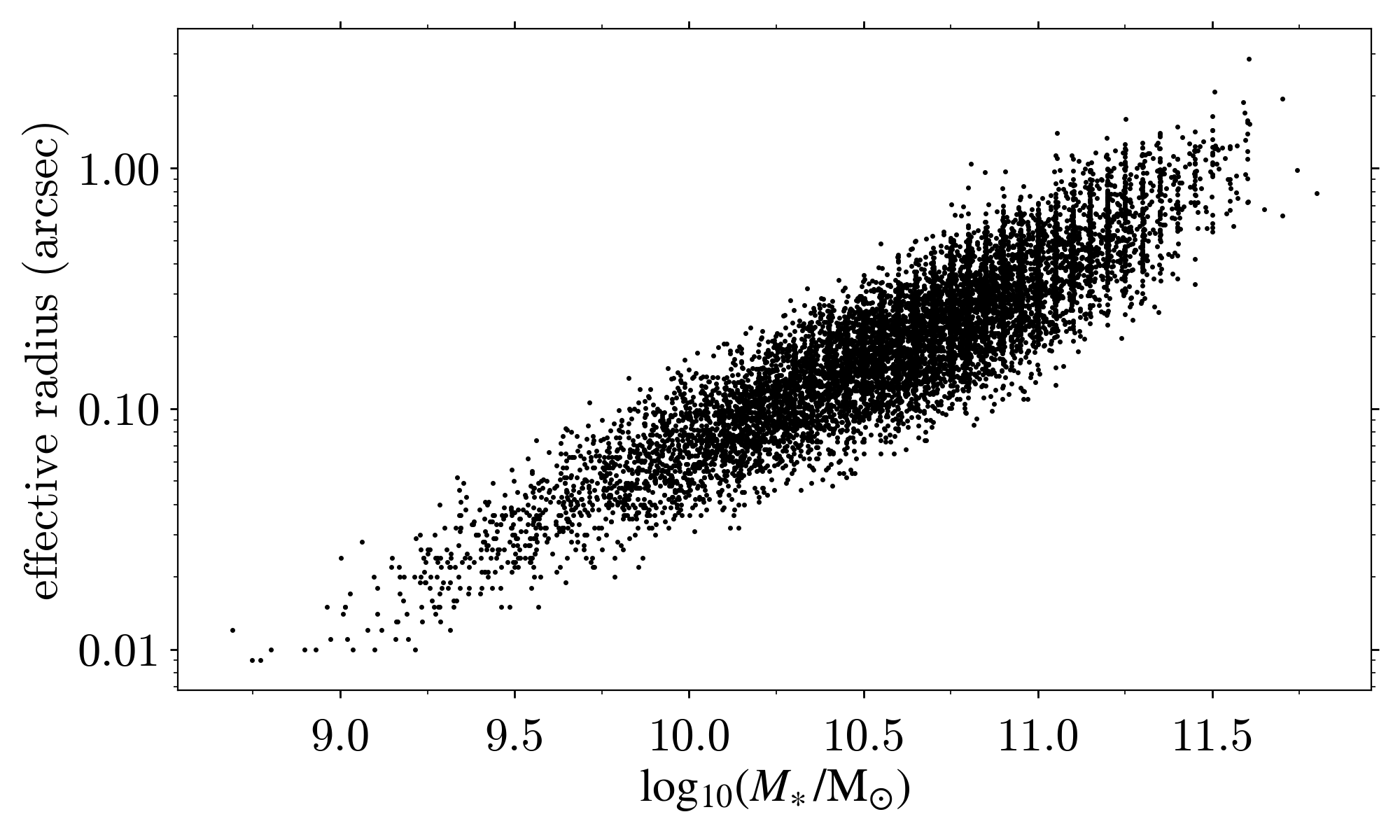}
	\caption{Distribution of sizes of quiescent COSMOS galaxies as a function of \Mstar,
		based on the relation and scatter measured by \citet{Wel2014}.}
	\label{fig:VdW}
\end{figure}

\subsubsection{Early-type galaxies}
\label{sec:mg}
Early-type galaxies over the redshift range \(z=\myrange{1}{2}\) have very red $zOYJH$ colours, 
that resemble the colours of high-redshift quasars at low S/N. 
There is a steep correlation between size and stellar mass for this population \citep[][]{Wel2014}. 
As a consequence, faint $J>22$ early-type galaxies at these redshifts will be very compact.
The \ang{;;0.3} pixel size of the \euc NISP instrument means that the surface brightness profiles 
of these faint early type galaxies will be poorly sampled, 
and therefore they may be mistaken for point sources, 
and classified as quasars. 
We now consider this possibility. 
While the pixel size of the \euc VIS instrument, $\ang{;;0.1}$, is much better, 
the detection \sn in the \(O\) band will be very low,
e.g., for a \(J=23\) early-type galaxy observed at \(z=1.5\),
the model \(O-J\) colour (described below) is greater than 2.5, 
implying \(\left(\sn\right)_O < 5\).

The best sample for investigating this issue, i.e.,
from the survey with the largest available area and that is deep enough for the \euc analysis, 
is the COSMOS sample presented by \citet{Laigle2016}. 
There is a total of \(1.38\,\dsq\) of crossover between COSMOS, and the NIR UltraVISTA bands.
We use only the quiescent objects from \citet{Laigle2016},
which are selected on the basis of a rest frame NUV/optical optical/NIR colour cut,
and we limit the analysis to redshifts \(z=\myrange{1}{2}\).
We additionally impose a magnitude cut, requiring \(J<24\),
which is sufficiently faint for the \euc wide survey.
(The quoted \(2\arcsecond, 3\,\sigma\) COSMOS depth is \(J = 25.2\),
so incompleteness at \(J<24\) will be negligible.)

All the COSMOS sources have a measured total $J$ magnitude,
an estimate of the total stellar mass (\Mstar), and a photometric redshift. 
To establish the distribution of sizes, 
we use the relations between effective radius 
(of the assumed de Vaucouleurs \(r^{1/4}\) profile) and stellar mass for quiescent galaxies, 
in different redshift bins, presented by \citet{Wel2014}. 
For a COSMOS galaxy with a particular stellar mass and redshift, 
we draw a random size from the distribution, given the specified variance. 
The resulting distribution of sizes of the COSMOS sample 
is plotted as a function of \Mstar in Fig.~\ref{fig:VdW}. 
Because we have a total magnitude for each source, 
we now have a sample that represents the complete magnitude/size/redshift distribution of the population at \(1<z<2\). 

At this point, ideally, we would simulate the detection, 
classification (star/galaxy discrimination), 
and photometry processes of the \euc pipeline on this sample, 
to derive the surface density of the population of early-type galaxies with $1<z<2$, 
classified as point sources, as a function of point-source magnitude and redshift. 
This detailed modelling has not yet been undertaken. 
Therefore to make progress, we start with the simplifying assumption 
that aperture photometry in a \ang{;;1.0} aperture 
provides a reasonable approximation of the \euc point-source photometry, 
recalling the large pixel size of the NISP instrument.

\begin{figure}[t] 
	\centering
	\includegraphics[width=8cm]{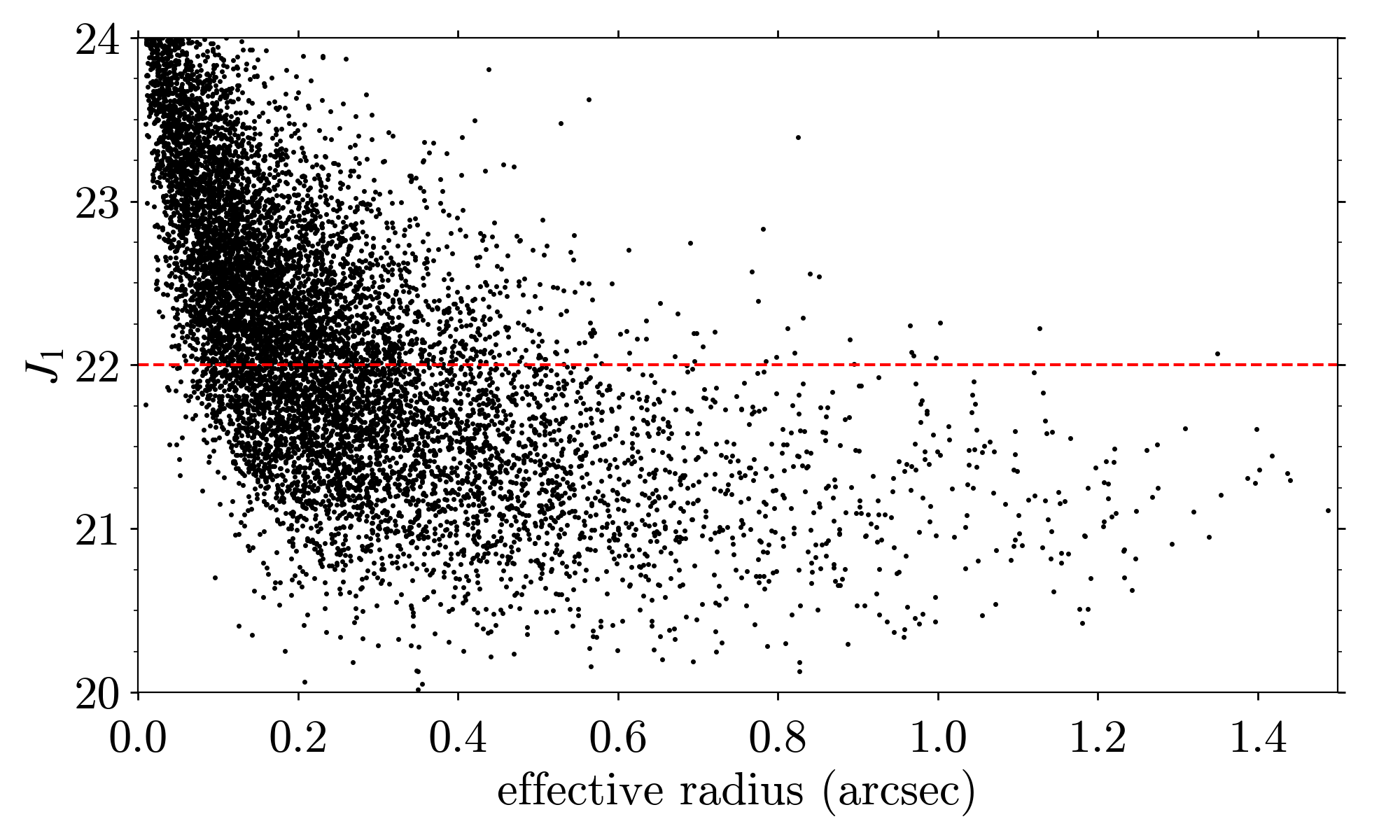}
	\caption{\ang{;;1} \(J\)-band aperture magnitude of COSMOS galaxies,
		as a function of half-light radius (see Fig.~\ref{fig:VdW}).
		Magnitudes fainter than \(J=22\), above the red dashed line,
		 are of particular relevance
		due to their predicted small radii.}
	\label{fig:Jreff}
\end{figure}

For each source in the COSMOS sample, 
we have integrated the \(r^{1/4}\) profiles to correct the total magnitudes to this aperture size. 
The resulting \ang{;;1} \(J\)-band magnitudes (denoted \(J_1\)) 
are plotted as a function of effective radius in Fig.~\ref{fig:Jreff}. 
The question now is what fraction of these galaxies would be classified as point sources? 
Using the BMC algorithm (for any sensible prior), we find that brighter than aperture magnitude $J_1=22$, 
the S/N is sufficiently high that 
quasars are cleanly discriminated from galaxies on the basis of their colours (Sect.~\ref{sec:qsf}).
The question of point/extended source discrimination is therefore immaterial at these brighter magnitudes.
Fainter than $J_1=22$ the colour discrimination begins to fall below $100\%$ success,
i.e., some quasars do not satisfy the selection threshold, and are misclassified.
Consequently we focus now on these fainter magnitudes.

\begin{figure}[t] 
	\centering
	\includegraphics[width=8cm]{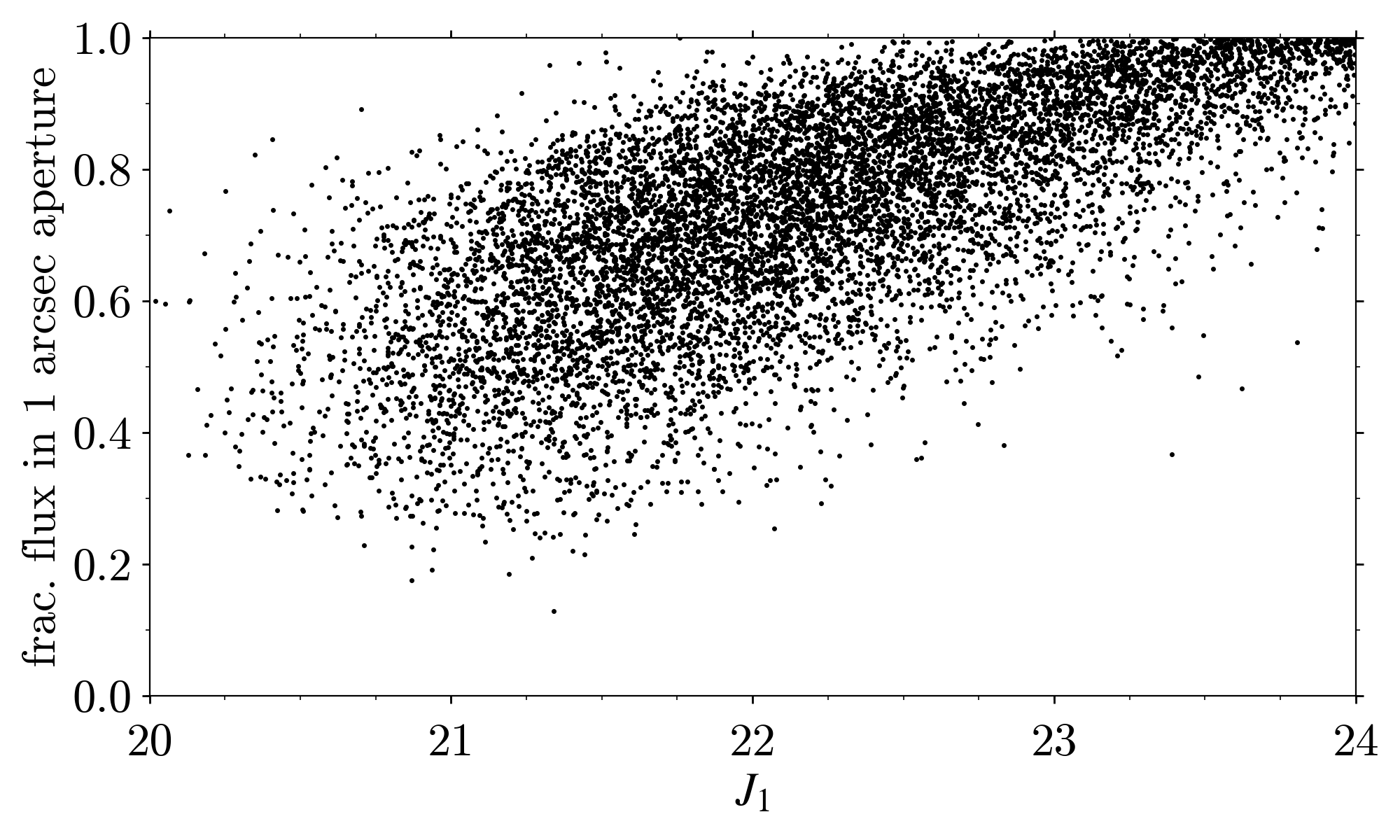}
	\caption{The fraction of flux contained in a \ang{;;1} diameter aperture,
		against \ang{;;1} \(J\)-band magnitude, measured for the COSMOS sample.
		The flux fractions are determined by integrating de Vaucouleurs profiles.}
	\label{fig:fluxcorr}
\end{figure}

As can be seen from Fig.~\ref{fig:Jreff},
fainter than $J_1=22$ the galaxies are very compact,
with effective radii mostly less than \ang{;;0.2},
meaning that many galaxies may be classified as point sources. 
Another way of seeing the problem is illustrated in Fig.~\ref{fig:fluxcorr}
which plots the fraction of the total flux inside the \ang{;;1} diameter aperture.
At $J_1=22$, the aperture contains on average $\sim70\%$ of the total flux,
increasing to $\sim90\%$ at $J_1=23$.
It is likely that most of these fainter galaxies,
detected in \(J\) at \(\sn \sim 10\), 
will be classified as point sources.
For the purposes of this paper,
we take a conservative approach and assume that all \(J_1>22\) early-type galaxies 
$1<z<2$ will be classified as point sources by \euc. 
We examine the consequences of this choice in Sect.~\ref{sec:discuss}.

To model the colours, we estimate formation redshifts ($\zf$) by combining redshift and age data provided by \citet{Laigle2016}. The histogram of $\zf$ shows a peak near $\zf=3$, with an extended tail towards higher redshifts. 
Consequently, we approximate the catalogue as two populations 
with a fraction 0.8 with \(\zf = 3\), and a fraction 0.2 with \(\zf = 10\), 
to try and encapsulate the range of formation redshifts seen in the data.
We compute colours for both formation redshifts from the evolutionary models of \citet[][]{Bruzual2003}.
The models are computed using the \citet{Chabrier2003} initial mass function and stellar evolutionary tracks prescribed by \emph{Padova 1994} \citep[e.g.][]{Girardi1996}. 
	We use single stellar populations with solar metallicity (M62; \(Z = 0.02\)) at our chosen formation redshifts and evolve them in time steps corresponding to \(\delta z = 0.1\) to cover the redshift range \(1.0 \leq z \leq 2.0\).

The galaxy surface density function is determined from a maximum likelihood fit
to the COSMOS data,
in terms of \(J_1\) and source redshift.
The functional form of the galaxy surface density function
in units of \(\mathrm{mag}^{-1}\,\mathrm{deg}^{-2}\) per unit redshift is
\begin{equation}
\begin{aligned}
&\Sigma(J_1,z) = 
\alpha\,
\exp\left\lbrace-\frac{1}{2}\left[\frac{J_1 - f\left(z\right)}{\sigma}\right]^2\right\rbrace\,
\exp\left[-\left(\frac{z-0.8}{z_0}\right)\right]\\
&f\left(z\right) = J_0 + b\,z,
\end{aligned}
\end{equation}
where we find the best-fitting parameters to be 
\(\left(\alpha,\sigma,J_0,b,z_0\right) = \left(8969,0.770,20.692,1.332,0.424\right)\).
We show this surface density function integrated over \(1.0 \le z \le 2.0\) 
	as a function of \(J\) magnitude in Fig.~\ref{fig:surfacedensity}.
We assume the same function is applicable to early-type galaxies with either formation redshift,
and scale the resulting weights by 0.8 for \(\zf=3\) and 0.2 for \(\zf=10\)
to reflect the distribution of \(\zf\) values seen in the COSMOS data.

Contamination from faint early-type galaxies may ultimately prove to be less
important for \(z>7\) quasar searches than presented in this paper. 
It may prove possible to increase the image
sampling in NISP, by drizzling multiple exposures of the same field, which will improve star/galaxy discrimination, 
and hence reduce the number of contaminants. 
Whether or not these objects will actually be classified as point sources
could be determined by including the population in future generations of \euc simulations.
Even if the effective radii of the galaxies are \ang{;;0.2} or smaller, 
it may be possible to identify light extending outside this radius 
if the PSF is well understood \citep[e.g.][]{Trujillo2006}.
Additionally, we have been somewhat conservative in assuming a minimum formation redshift of $z=3$, and a single burst of star formation. The COSMOS sample includes galaxies with later formation redshifts, which may also have some ongoing star formation,
rendering them more visible in the $O$ band \citep{Conselice2011}. 

 
\section{Results}
\label{sec:qsf}

\begin{figure*}[t]
	\centering
	\subfloat[\euc data only.]{%
		\includegraphics[width=9cm]{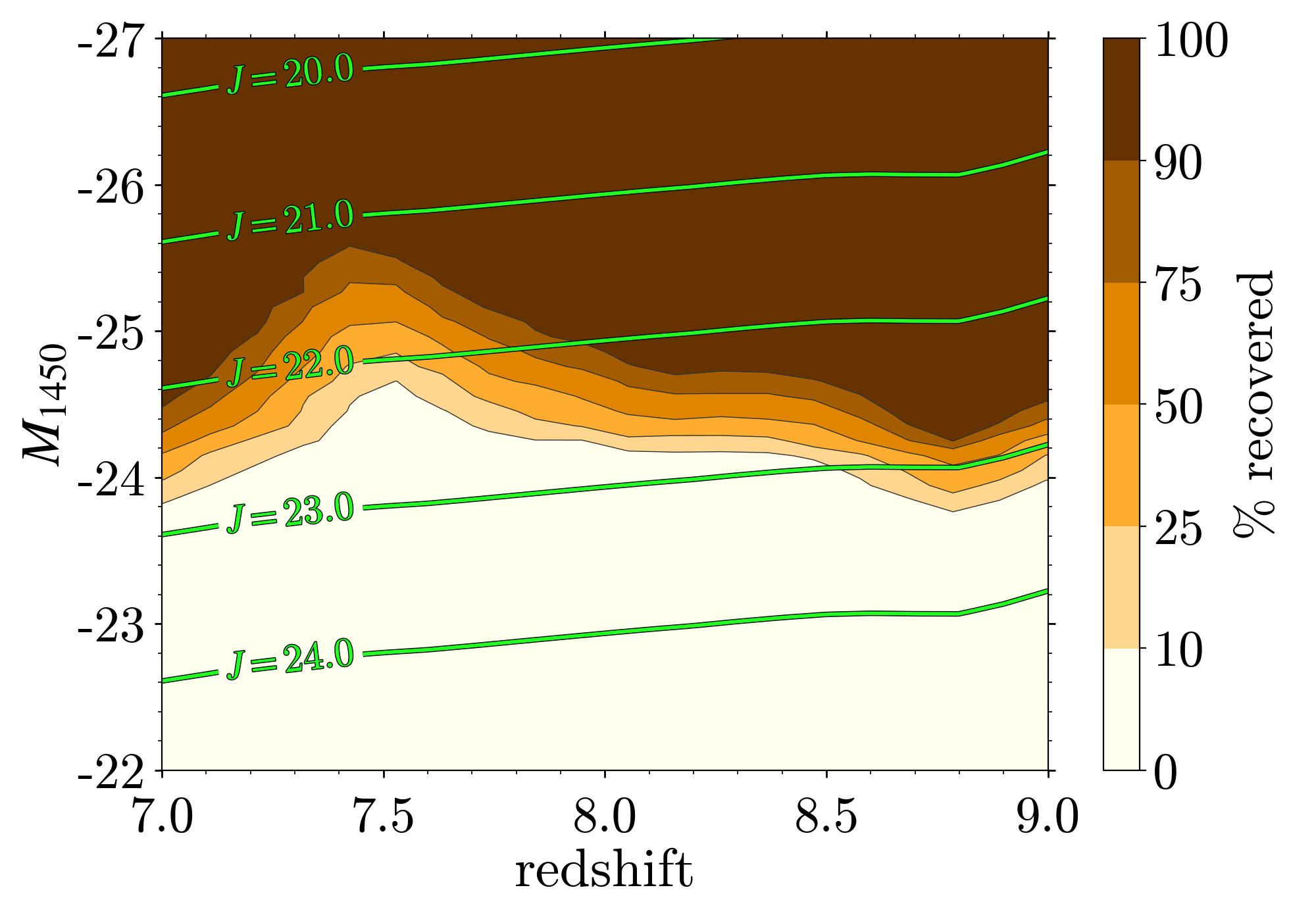}%
		\label{fig:qsfng}%
	}\,
	\subfloat[\euc \(O\) band replaced with ground-based optical data.]{%
		\includegraphics[width=9cm]{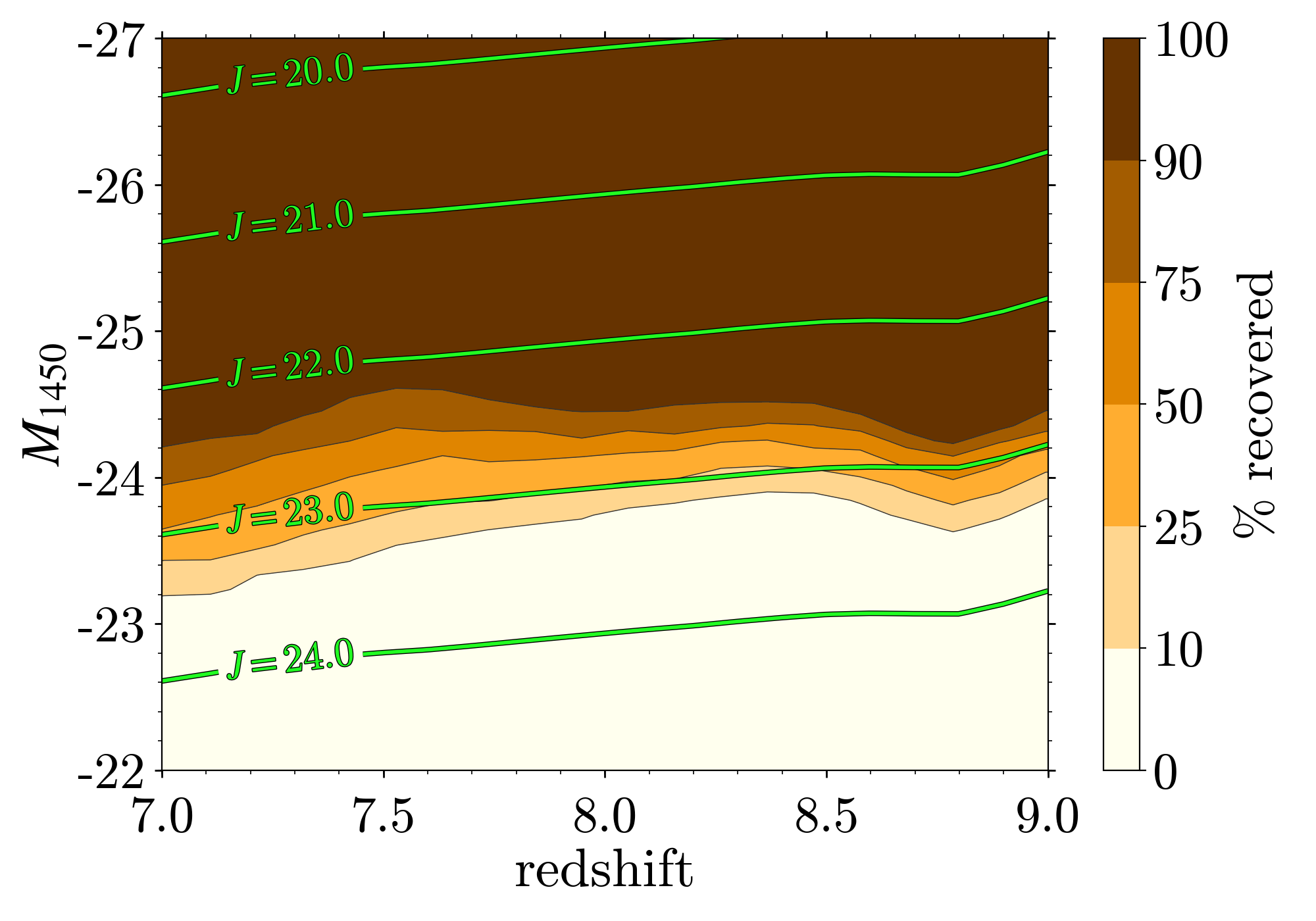}%
		\label{fig:qsfg}%
	}
	\caption{Quasar selection functions determined using the BMC method
		for \euc $YJH$ data with a) \euc optical data, b) ground-based optical data.
		A quasar is defined as selected if \(P_{\mathrm q} > 0.1\).
		Contours of apparent magnitude are indicated by the labelled green lines.}
	\label{fig:qsfbmc}
\end{figure*}

To model \euc selection of high-redshift quasars, 
we apply the BMC and minimum-\(\chisq\) model fitting methods outlined in Sect.~\ref{sec:methods}
to the simulated quasar grids. 
The main results are selection functions (Figs.~\ref{fig:qsfbmc} and \ref{fig:bmccsq}),
which we combine with population models to obtain predicted numbers (Table~\ref{tab:counts}).
In Sect.~\ref{sec:bmcresults} we discuss the results from the BMC technique,
and consider the impact of ground-based optical data.
In Sect.~\ref{sec:csqresults} we compare against the \(\chisq\) method.
In Sect.~\ref{sec:contam} we consider the extent of contamination 
by MLTs and ellipticals which are selected as quasar candidates.

\begin{figure*}[t]
	\centering
	\subfloat[Predicted yield
	in redshift bins of \(\Delta z = 0.1\).]{%
		\includegraphics[width=9cm]{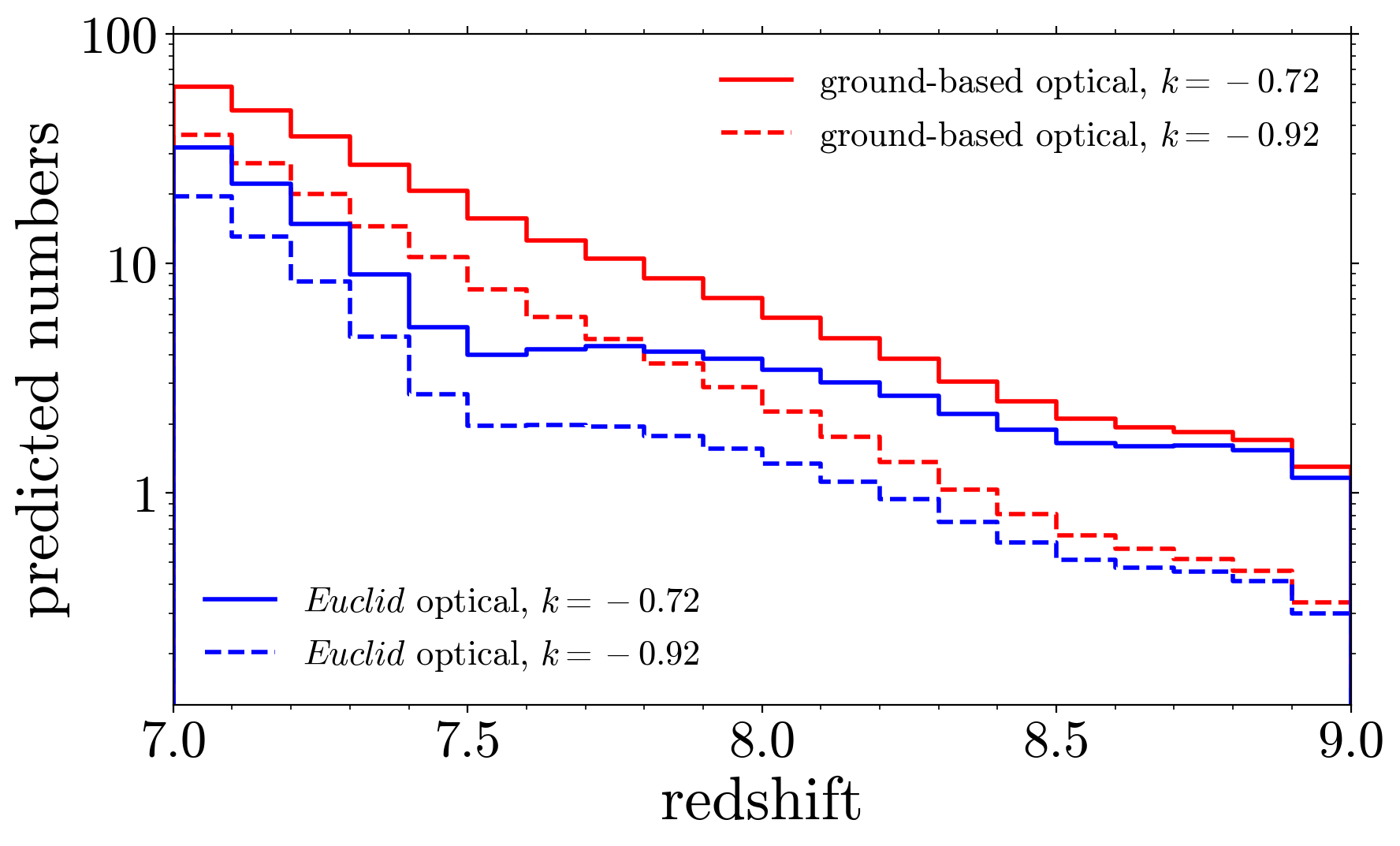}%
		\label{fig:countsz}%
	}\,
	\subfloat[Cumulative predicted yield 
	as a function of \(J\) magnitude.]{%
		\includegraphics[width=9cm]{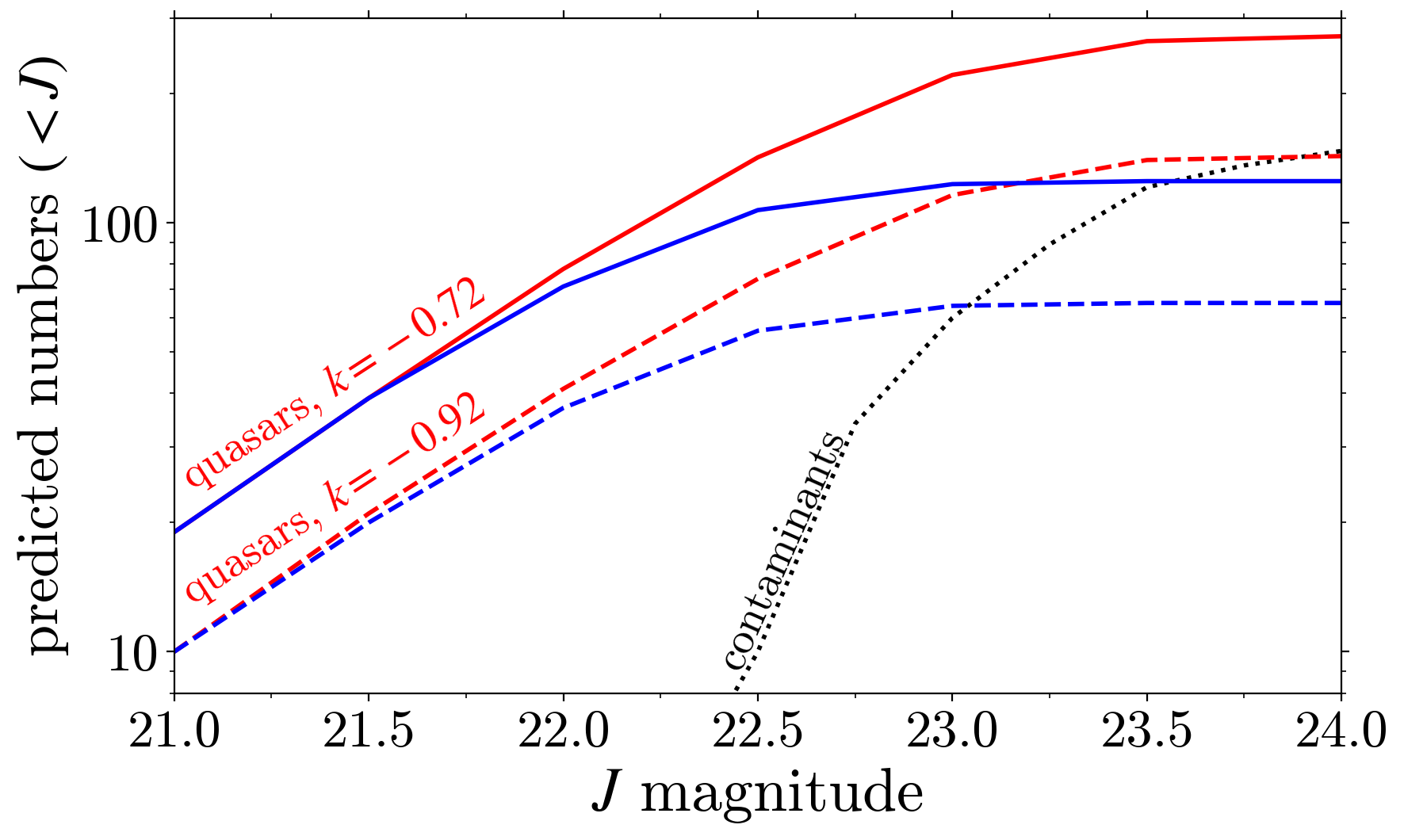}%
		\label{fig:countsJ}%
	}
	\caption{Predicted numbers of \(7<z<9\) quasars as a function of redshift (left) and \(J\) magnitude (right),
		determined by integrating the QLF over the selection functions presented in Fig.~\ref{fig:qsfbmc}, 
		and assuming an area of \(15\,000\,\dsq\). 
		Blue: \euc data only. Red: \euc \(O\) band replaced with ground-based optical data. 
		Solid lines $k=-0.72$. Dashed lines $k=-0.92$. 
	The additional black dotted curve on the right-hand panel indicates the estimated number of contaminants 
	selected as quasar candidates as a function of magnitude, 
	assuming ground-based optical data, and so should be compared to the red curves, labelled.}
	\label{fig:counts}
\end{figure*}

\begin{table}
	\centering
	\advance\leftskip-3cm
	\advance\rightskip-3cm
	\caption{Summary of predicted numbers of \euc wide survey quasars in redshift bins,
		determined by integrating the QLF over the BMC and minimum-$\chi^2$ quasar selection functions.
		Results are presented incorporating either \euc or ground-based optical data,
		for two redshift evolutions of the QLF.
		Numbers from the minimum-\(\chisq\) model fitting are additionally given in brackets.} 
	\begin{adjustbox}{width=0.5\textwidth,totalheight=\textheight,keepaspectratio}
		\label{tab:counts}
		\begin{tabular}{ccccc}
			\hline \hline \\[-2ex]
			Redshift range & \multicolumn{2}{c}{\euc optical} & \multicolumn{2}{c}{Ground-based optical} \\
			& \(k = -0.72\) &   \(k = -0.92\)  & \(k = -0.72\) &   \(k = -0.92\)          \\
			\hline \\[-2ex]
			\(7.0< z <7.5\)&    87 (41)    &     51 (24)      &    204 (91)  &        117 (52)           \\ 
			\(7.5< z <8.0\)&    20 (13)    &      9 (6)       &    45 (26)    &        19 (11)           \\
			\(8.0< z <8.5\)&    11 (11)    &      4 (4)       &    16 (14)    &         6 (5)               \\
			\(8.5< z <9.0\)&     6 (6)     &      2 (2)       &    7  (7)     &         2 (2)               \\
			\hline		
		\end{tabular}
	\end{adjustbox}	
\end{table}

\subsection{Bayesian model comparison}
\label{sec:bmcresults}

We present the quasar selection function determined with the BMC technique,
and using \euc optical data, in Fig.~\ref{fig:qsfng}.
This shows that over the redshift range $8<z<9$, 
quasars may be selected fainter than the previously assumed limit of $J=22$.
The situation is worse over the redshift range $7<z<8$, 
where the discrimination against MLT dwarfs is relatively poor, 
and the typical depth reached is $J\sim 22$. 

The selection function for the case where deep ground-based $z$-band data are available is presented in 
Fig.~\ref{fig:qsfg}. There is only a small difference between the individual LSST and Pan-STARRS selection functions,
driven by the different depths of the surveys.
For simplicity, we combine the LSST and Pan-STARRS selection functions in the ratio 2:1 
(to reflect the respective areas),
and present a single `ground-based' selection function. 
As can be seen, the use of $z$-band optical data, compared to \euc $O$-band data,
	means the quasar survey can reach up to 1 mag deeper over the redshift range $7<z<8$.
	There is also improvement over the redshift range $8<z<8.5$,
	while between $8.5<z<9$ the improvement is smaller.
    Broadly speaking, we now recover quasars as faint as \(J\sim23\) over the full redshift range $7<z<9$.	 

At redshifts $7<z<8$ the survey depth is set by the ability to discriminate against MLT dwarfs. Over the redshift range $8<z<9$
the contaminant weights are more balanced,
i.e., the quasars that are not recovered
are misclassified either as early-type galaxies or MLTs.
In Sect.~\ref{sec:discuss} we discuss the individual impact 
of the two contaminating populations on the quasar selection functions.

To estimate the number of quasars that can be detected in the \euc wide survey,
we integrate two different QLFs over the selection functions 
We adopt the \citet{Jiang2016} $z=6$ QLF\footnote{
We note that the range of quasar yields presented in this paper 
		determined by integrating the \citet{Jiang2016} QLF 
		over our selection functions, for the two chosen values of \(k\), encompasses the results obtained
		in the case that the more recent \citet{Matsuoka2018c} QLF is used instead.}, 
with the decline towards higher redshift parametrised as $\Phi\propto10^{k(z-6)}$, 
and calculate numbers for two values of $k=-0.72$, and $-0.92$. 
The first value assumes that the rate of decline over the redshift interval $5<z<6$ 
measured by \citet{Jiang2016} continues to higher redshifts. 
The second value assumes that the decline continues to steepen with increasing redshift.
The value of \(k=-0.92\) is arbitrary,
and was chosen simply to present a more pessimistic forecast for comparison.
Relative to the case where \(k=-0.72\), using \(k=-0.92\) implies
the space density of quasars is a factor of 1.6 (2.5, 4) lower at \(z = 7~(8, 9)\).
We plot the predicted numbers in redshift bins in Fig.~\ref{fig:countsz}, 
for \euc optical data (blue), and ground-based optical data (red), 
for the two different assumed values of $k=-0.72, -0.92$ (solid, dashed respectively). 
The smaller numbers and steeper decline for $k=-0.92$ compared to $k=-0.72$ are easy to see. 
The benefit of using $z$-band data compared to the \euc $O$ band is also very clear, 
with the largest improvement near $z\sim7.5$, 
and an average improvement in numbers by a factor of \(\sim2.3\), detected over the range $7<z<8$.
The cumulative numbers $7<z<9$ are plotted
as a function of \(J\)-band magnitude in Fig.~\ref{fig:countsJ}.
We summarise the total predicted yield
in redshift intervals $\Delta z=0.5$ in Table~\ref{tab:counts}.
	The counts in Table~\ref{tab:counts} are evaluated down to the assumed \euc wide survey limit.
	Assuming $z$-band data are available,
	Fig.~\ref{fig:countsJ} implies the majority of \(z = \myrange{7}{9}\) quasars detected in \euc will be brighter than \(J=23\).
	However, despite the relatively poor selection efficiency at fainter magnitudes,
	we predict \euc can detect up to 50 quasars with \(J>23\) (assuming \(k=-0.72\)),
	which will be in the range \(z = \myrange{7}{8}\) where the space density is highest.

The predicted number counts considerably exceed those from \citet{Manti2017},
although the uncertainty in their calculation spans two orders of magnitude,
and refers to a brighter sample (they use a \(10\,\sigma\) limit for \euc).
That work predicts, for example, two \(7<z<8\) quasars from the \euc wide survey 
using a double power law parametrisation of the QLF,
or 20 sources using a Schechter function parametrisation,
but also underestimates the actual yields from VIKING and the UKIDSS LAS at lower redshift.

The predicted yield of $z>8$ quasars presented in \cite{Laureijs2011} 
used a considerably more gentle rate of decline, $k=-0.47$, 
and assumed a detection limit of $J=22$. 
The computed numbers presented in Table~\ref{tab:counts} 
show that the improvement in selection method, 
using the BMC method rather than colour cuts, 
and so reaching deeper, 
offsets to a large extent the steeper rate of decline of the quasar space density 
now believed to exist at high redshifts. 
\cite{Laureijs2011} predicted \euc would find 30 quasars $J<22$, $z>8.1$. 
For $k=-0.72$, using BMC, we predict 23 quasars $z>8.0$. 
Even if the rate of decline is as steep as $k=-0.92$ 
we predict that \euc can find 8 quasars with redshifts $z>8.0$. 
In 2011 it was expected that substantial samples of $7<z<8$ quasars 
would exist by the time of the launch of \euc. 
This expectation has changed in the interim. 
There are currently five $z>7$ quasars known, 
and prospects for increasing this number much before \euc is launched are poor. 
With \euc we expect to detect over one hundred
quasars brighter than \(J=23\) with redshifts \(7<z<8\),
even if the redshift evolution of the QLF is as steep as $k=-0.92$.

\subsection{$\chisq$ model fitting}
\label{sec:csqresults}

\begin{figure}[t]
	\centering
	\includegraphics[width=9cm]{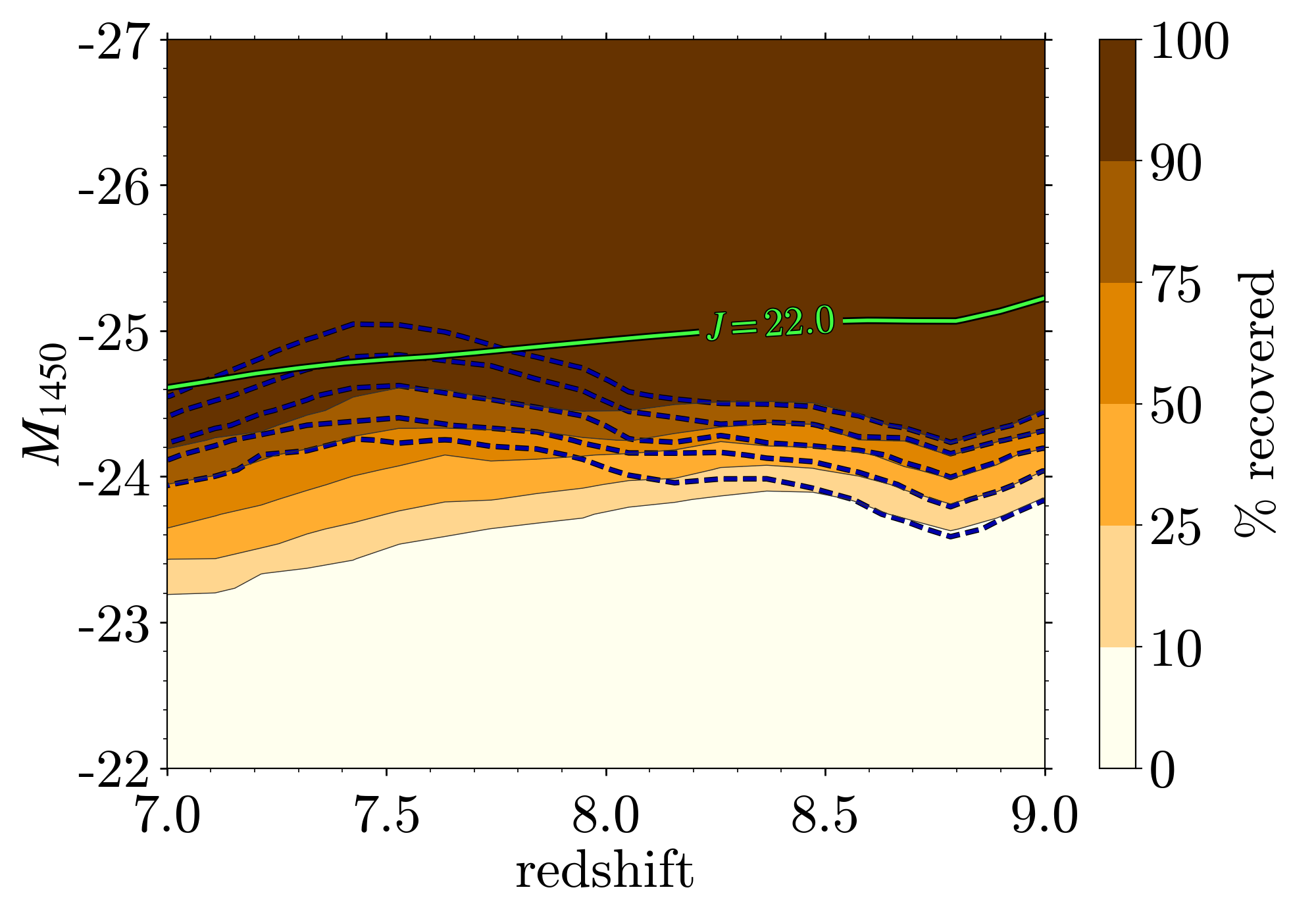}
	\caption{Quasar selection functions determined using BMC 
		(filled contours; same as Fig.~\ref{fig:qsfg}) 
		and $\chisq$ model fitting (dashes),
	assuming \(z\)-band data are available.
		Contour intervals are the same in both cases.
	}
	\label{fig:bmccsq}
\end{figure}

Following the procedure described in Sect.~\ref{sec:chioverview}, 
we additionally measure quasar selection
using minimum-\(\chisq\) model fitting.
We integrate QLFs over these new selection functions,
and present the resulting numbers in redshift bins in Table~\ref{tab:counts}.
Taken as a whole, the BMC method significantly outperforms the $\chisq$ model fitting.
In general, the inclusion of surface density information for each population
    improves the depth to which one is able to select high-redshift quasars \citep{Mortlock2012}. 
 
However, as seen in the selection functions in Fig.~\ref{fig:bmccsq},
the differences in the contours depend on redshift.
At lower redshifts $7<z<8$
the BMC contours in Fig.~\ref{fig:bmccsq} are 
around \(0.5\)\,mag. deeper than the contours of the $\chisq$ method, 
and the yield is around a factor of two greater (Table~\ref{tab:counts}).
By contrast, at \(z\gtrsim8.2\) the methods apparently perform equally well,
as quasars are easily separated from other populations on the basis of \(Y-J\),
meaning contaminant models are always 
poor fits to the simulated photometry.
As such, the predicted quasar yield is very similar for both methods between \(z=\myrange{8.0}{8.5}\),
and at \(z>8.5\) the predicted numbers are the same using either method.
In Sect.~\ref{sec:contam} we explore relaxing the selection cuts 
of the \(\chisq\) method to produce a deeper sample.
However, doing so (while keeping contamination low) 
does not significantly improve quasar selection over \(7.0<z<8.0\),
meaning the predicted number counts remain lower than for the BMC method.
We conclude that the absence of prior population information in the $\chisq$ method
places a limit of \(J\sim22.5\) on the quasars \(7.0<z<8.0\) that can be detected in \euc in this way.


\subsection{Sample contamination}
\label{sec:contam}

An important further consideration is the selection of contaminants in the \euc wide survey,
i.e., the number of MLTs and early-type galaxies that pass our selection criteria.
We simulate a realistic number of contaminants over the full wide survey area,
with magnitudes (both populations) and redshifts (ellipticals only)
drawn from the surface density functions described in Sect.~\ref{sec:pops}.
The sources that we generate have a true \(J\) magnitude 
up to one magnitude fainter than the survey limit,
to allow sources to scatter bright when we simulate the noisy photometry.
In total, we simulate \(8.6\times10^7\) MLTs, and \(4.8\times10^7\) ellipticals with \(1<z<2\).
To make the \pq calculation more manageable, 
and to focus on the sources that are most likely to be of interest,
we take a cut on \(\chisq\), 
discarding all sources that are reasonably well fit by any contaminant template SED
(\(\chisq_{{\rm red, c(best)}}<5\)).
The remaining sample contains \(6.1\times10^5\) MLTs, and \(1.5\times10^5\) galaxies.
We apply the BMC method to these samples, assuming that ground-based data are available.
In total, we recover 147 sources with \(P_{\mathrm q} > 0.1\).
The majority (126) are brown dwarfs, with 21 galaxies additionally recovered.
The dwarf stars have spectral types between \myrange{M9}{T7},
although more than 75\% of these sources are in the range \myrange{T2}{T4}.
Later T-types are much bluer than quasars in \(J-H\), 
which is typically sufficient to discriminate between the two populations.
The \(z-Y\) colour of the recovered sources has typically scattered very red, such that \(z-Y>3\),
making the measured SED of each object a close match to the quasar templates (Fig.~\ref{fig:colourcolour}).

We show the cumulative contaminant numbers using the BMC method as a function of \(J\) magnitude 
as the black dotted line in Fig.~\ref{fig:countsJ}.
This prediction should be compared to the red curves, 
which are also based on the availability of ground-based \(z\)-band data.
Brighter than \(J=22.5\) the number of recovered contaminants is very low,
suggesting quasars will be very efficiently recovered at these magnitudes.
However, as the \sn falls further, the contamination starts to increase.
Using BMC, the majority of quasars detected by \euc will be brighter than \(J=23\).
At this magnitude limit, the implied selection efficiency, 
defined as the ratio of quasars to the total number of selected sources,
will be around two thirds, depending on the exact QLF evolution.
By \(J=24\), the number of contaminants is comparable to the predicted quasar numbers for \(k=-0.92\),
implying a selection efficiency of around a half.

In the case where only \euc optical data are available,
	the number of contaminants is seen to fall, 
	e.g., we now select 50 brown dwarfs with spectral types \myrange{M9}{T7}
	with \(P_{\mathrm q} > 0.1\).
	The numbers are similar to the case where the \(z-Y\) colour is used as faint as \(J\sim22.7\);
	however, few sources are selected fainter than this limit 
	when using the $O$ band in the BMC calculation rather than $z$. 
	This is a similar picture to the quasar numbers as a function of magnitude presented in Fig.~\ref{fig:countsJ}. 
	As discussed in Sect.~\ref{sec:ground}, 
	the $z-Y$ colour straddles the the \lya break more closely than $O-Y$, 
	improving the contrast between populations,
	and so enhancing our sensitivity to quasars. 
	This allows the search for quasars to go deeper, 
	at the expense of additional contamination.

Repeating the analysis (with \(z\)-band data) for the \(\chisq\) method, 
the total number of contaminants is 25 using the cuts in Sect.~\ref{sec:chioverview},
implying a high selection efficiency at all magnitudes
(although the BMC method is comparably efficient brighter than \(J=22.5\),
the limit of the \(\chisq\) method discussed in Sect.~\ref{sec:csqresults}).
This might suggest that the \(\chisq\) cuts could also be relaxed,
to allow a deeper search for quasars in \euc, but a preliminary analysis indicates this is not fruitful.
As a test we re-calculated the predicted numbers of quasars and contaminants with slightly looser cuts:
\(\chisq_{{\rm red,c(best)}} > 9\); and \(\chisq_{{\rm red,c(best)}}/\chisq_{{\rm red,q(best)}} > 2.5\).
Doing so increased the contamination by factor of three,
while the number of quasars increased only by around \(10\%\), driven by a very small improvement over \(7<z<8\).
In summary the \(\chisq\) method has similar effectiveness to the BMC method over the magnitude and redshift range over which it is sensitive, Fig.~\ref{fig:bmccsq}, but the BMC method results in much higher predicted numbers of quasars because it reaches 0.5 mag. deeper over the redshift range \(7<z<8\). 

The observed decline in selection efficiency with apparent magnitude will have a bearing on future follow-up strategy, 
which we discuss further in Sect.~\ref{sec:followup}.
However, follow-up will prioritise the highest probability candidates, 
which will typically be the brightest. 
If future follow-up resources are limited then, 
e.g., a magnitude cut can be applied to ensure a complete sample and allow measurements of the QLF.

\section{Discussion}
\label{sec:discuss}

The quasar yield predicted in Sect.~\ref{sec:qsf} and summarised in Table~\ref{tab:counts} confirms that 
\euc can make a major contribution to EoR science in the 2020s.
We now explore some of the implications of the simulation work presented in this paper. 
In Sect.~\ref{sec:timeline} we discuss the likely timeline for quasar discoveries
with \euc. 
In Sect.~\ref{sec:qlfconstrain} we consider the extent to which \euc can constrain the QLF.
In Sect.~\ref{sec:followup} we explore some of the challenges
in terms of the follow-up of \euc high-redshift quasar candidates.

We additionally examine some of the uncertainties that have a bearing on the calculation presented in this work.
In Sect.~\ref{sec:cosmosdiscuss} we use COSMOS data to further investigate
our choice of contaminating populations in this work.
In Sect.~\ref{sec:elldiscuss} we consider to what extent the assumptions 
made about the early-type galaxy population influence the predicted numbers.
In Sect.~\ref{sec:lx_cx} we explore the extent to which quasar selection using \euc 
is affected by the range in quasar properties.
We find that neither of these uncertainties is important, 
and the dominant uncertainty in the calculations presented here is the value of the parameter $k$.

\subsection{Potential status mid-2020's}
\label{sec:timeline}

\begin{table}[t] 
	\centering
	\advance\leftskip-3cm
	\advance\rightskip-3cm
	\caption{Potential quasar yield in redshift bins,
		following the \euc DR1 release planned for 2024.
		Numbers are determined over \(1250\,\dsq\) of the southern hemisphere,
		assuming LSST one-year data are available in the optical.
		Results are shown for two evolutions of the QLF.}
	\label{tab:counts2024}
	\begin{tabular}{ccc}
		\hline \hline \\[-2ex]
		Redshift range      &  \(k=-0.72\) &  \(k=-0.92\) \\ 
		\hline \\[-2ex]
		\(7.0< z <7.5\)  &  18.8        &  10.7        \\
		\(7.5< z <8.0\)  &  4.1         &  1.8         \\
		\(8.0< z <8.5\)  &   1.5        &  0.5         \\
		\(8.5< z <9.0\)  &   0.6        &  0.2         \\
		\hline \\[-2ex]
		Total               &  24.9        &  13.2        \\
		\hline		
	\end{tabular}
\end{table}

With the launch of \euc currently planned for mid 2022, 
and the full \(15\,000\,\dsq\) of wide survey not expected to be available
until some seven years after launch (\nth{3} data release; DR3),
it will be over a decade before the number count predictions
in this paper are fully realised.
Nevertheless, intervening data releases will offer opportunities 
to carry out excellent quasar science.
Importantly, the desired wide survey depth for each tile
is achieved in a single visit,
while the area is built up over time;
hence the selection functions are applicable to all \euc releases,
depending on the availability of complementary optical data.
The first \euc quick release (Q1)
is expected around 14 months after the start of survey operations,
and will only cover a small area,
with the exact size and location yet to be determined.
However, assuming \(k=-0.72\), and an area of \(50\,\dsq\),
the predicted yield, $7<z<9$, is at most one quasar,
even when \(z\)-band data are used.
Nevertheless, Q1 data will offer an opportunity
to test the proposed selection methods, 
and get a sense of the expected contamination
rate when applied to real data.

This assessment of the predicted quasar numbers from Q1
	has assumed that data from this initial release will match the wide survey depth, 
	i.e., \(YJH=24\).
	We have not considered the possibility that these fields form part of the \euc deep survey,
	which will ultimately go two magnitudes fainter than the wide survey in all bands
	However, as mentioned in Sect.~\ref{sec:euclid}, 
	deeper observations would not make a significant difference to the predicted numbers for Q1.
	In principle, the quasar yield would be maximised by
	surveying a wider area, rather than going deeper.
	The single visit depth of \(J=24\) means for \(z=\myrange{7}{9}\) quasars,
	we are already sampling the faint-end slope of the QLF 
	\citep[\(M_{1450} > -24.9\),][]{Matsuoka2018c}.
	Without any consideration of completeness,
	the \citet{Jiang2016} QLF integrated to \(J=24\) implies
	one \(z=\myrange{7}{9}\) quasar per \(20\,\dsq\).
	Going one magnitude fainter, this density increases to 
	one \(z=\myrange{7}{9}\) quasar per \(8\,\dsq\),
	but the additional depth would require six observations of the field to achieve.

Looking further ahead, \euc DR1, comprising the first year of survey data,
is anticipated in the second year after the nominal mission start (i.e., mid-2024).
DR1 should cover \(2500\,\dsq\) in total,
split equally between the northern- and southernmost sky.
In the northern hemisphere, Pan-STARRS is only expected 
to have reached a \(5\sigma\) depth of \(z=24.1\) by DR1;
hence, selection will be somewhat worse over \(z=\myrange{7}{8}\)
than assumed in Fig.~\ref{fig:qsfg}.
However, access to one-year LSST data is a realistic prospect.
In Table~\ref{tab:counts2024} we present predicted numbers 
for the southern hemisphere following DR1,
assuming LSST data are available, and assuming that an area of \(1250\,\dsq\) is covered by LSST.
Even with stronger redshift evolution, $k=-0.92$,
the quasar yield from DR1 will potentially be significant,
especially when combined with additional discoveries from the northern hemisphere.
We would anticipate considerably more than ten
sources over the redshift range \(7<z<9\),
which would potentially include the first discoveries at \(z>8\),
from the full DR1 area.
DR1 will therefore likely be an exciting prospect for high-redshift quasar science,
with scope for significant development with subsequent \euc data releases.

\subsection{Quasar luminosity function constraints}
\label{sec:qlfconstrain}

\begin{figure}[t]
	\centering
	\includegraphics[width=9cm]{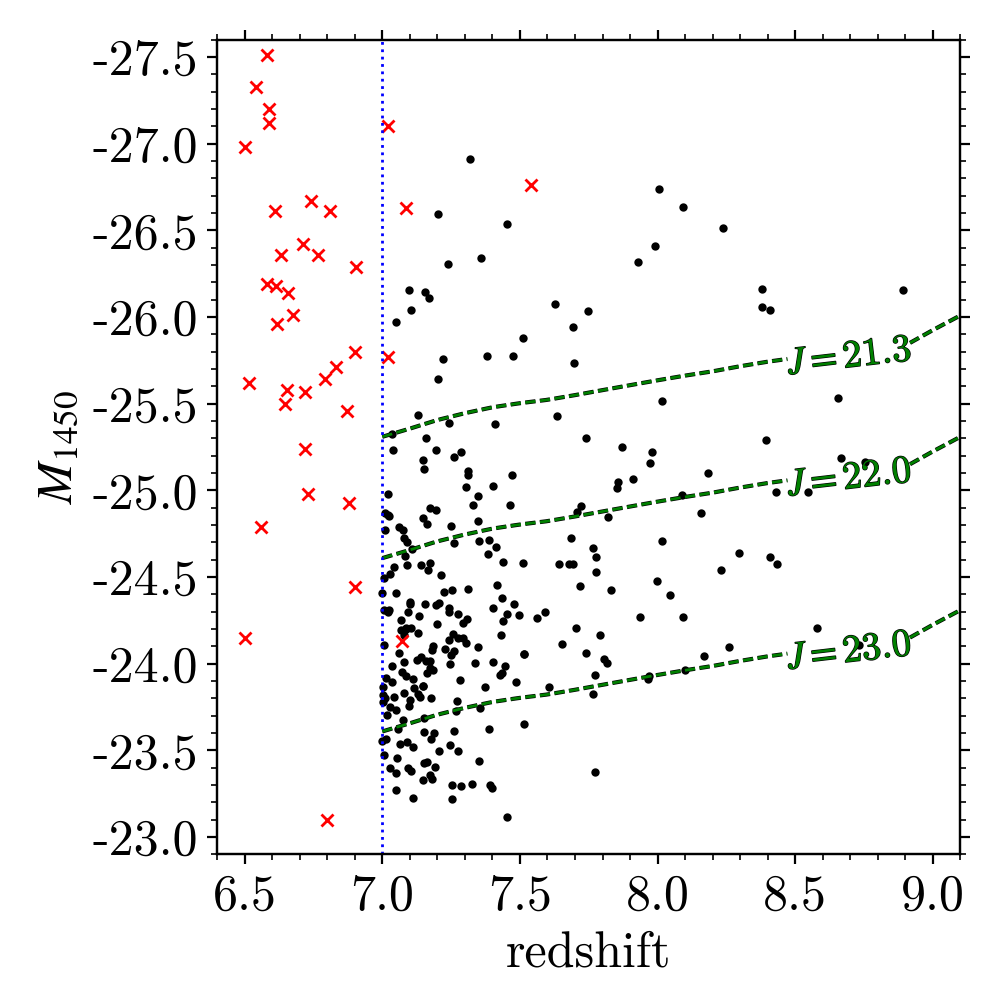}
	\caption{\(M_{1450}\,/\,z\) plane with all \(z>6.5\) quasars
		with published redshifts and luminosities at 1450\,\AA~(red crosses),
		and a simulated \euc wide survey quasar sample (black points),
		with random luminosities and redshifts
		drawn from the ground-based selection function (Fig.~\ref{fig:qsfg}).
		The blue dotted line indicates the redshift cut-off of this work.
		The green dashed contours indicate the apparent magnitudes 
		considered in Sect.~\ref{sec:followup},
		in the context of ground-based follow-up spectroscopy and contamination.
		Discovery papers for the known quasars are:
		\citet{Mortlock2011,Venemans2013,Venemans2015a,
			Matsuoka2016,Matsuoka2018a,Matsuoka2018b,Matsuoka2019,
			Mazzucchelli2017,Tang2017,Koptelova2017,Reed2017,
			Wang2017,Wang2018a,Wang2018b,Banados2018,Yang2019}.
	}
	\label{fig:Mzplane}
\end{figure}

	A sample of \euc \(z>7\) quasars will provide constraints on the QLF. 
	To illustrate the potential of \euc, we simulate a full wide survey quasar sample,
	assuming \(k=-0.72\) and that \(z\)-band data are available,
	with redshifts and magnitudes drawn from the distributions in Fig.~\ref{fig:counts}. 
	We plot the redshifts and luminosities of this simulated sample in Fig.~\ref{fig:Mzplane},
	shown alongside all \(z>6.5\) quasars which have been published to date 
	(references in caption). 

	Fig.~\ref{fig:Mzplane} illustrates the redshifts and luminosities
	at which \euc will have a particularly large impact. The \euc wide survey will be especially useful
	for measuring the redshift evolution of the quasar number density, 
	parametrised by \(k\).
	Previous works have used bright (\(M_{1450}<-26\)) quasars to determine \(k\) 
	\citep[e.g.][]{Fan2001a,McGreer2013,Jiang2016};
	however, by the time \euc data are available,
	the \(6.5<z<7\) QLF is likely to be sufficiently well determined
	to measure the evolution of number density at fainter magnitudes \citep[see, e.g.,][]{Matsuoka2018c}.
	As an illustrative example, 
	we consider the constraints that can be put on \(k\),
	assuming the number density has been well measured to a depth of \(M_{1450}=-25\) over $6.5<z<7.0$.
	The simulated \euc sample contains 24 \(7.5<z<8.5\) sources with \(M_{1450}<-25\).
	Assuming \(k=-0.72\) represents the true redshift evolution over \(z = \myrange{7}{8}\),
	the \euc sample implies we could measure \(k\) to a $1\sigma$ uncertainty of 0.07
	over that redshift range.

	\euc will also place strong constraints on the
	faint end of the quasar luminosity function.
	At $z=6$, the characteristic `knee' magnitude, \(M^*_{1450}\),
	was recently well constrained by \citet[][\(M^*_{1450} = -24.9^{+0.75}_{-0.90}\)\,mag]{Matsuoka2018c};
	the same authors also obtained a faint-end slope, \(\alpha=-1.23^{+0.44}_{-0.34}\).
	As illustrated in Fig. \ref{fig:Mzplane}, \euc will produce a large sample of quasars $7<z<8$ fainter than $M_{1450}=-25$, 
	which will allow the faint-end slope to be measured precisely over this redshift range,
	and also allow the evolution of the break from \(z\sim6\) to \(z>7\) to be determined.

\subsection{Follow-up demands}
\label{sec:followup}

	Confirmation and exploitation of the high-redshift quasar candidates identified by \euc will require follow-up spectra,
	e.g., to measure the redshifts, and to study the \lya damping wing to measure the cosmic density of neutral hydrogen.
	We use the \euc selection function to determine predicted numbers as a function of \(J\),
	as in Fig.~\ref{fig:countsJ}, but for the separate ranges \(z = \myrange{7}{8}\), and \(z = \myrange{8}{9}\).
	In both redshift bins, the median magnitude is \(J\sim22.5\),
	while the \nth{10} and \nth{90} percentiles are respectively \(J\sim21.3\) and \(J\sim23.3\),
	which we take to be indicative of the typical range of depths that we would need to reach.
	For a particular high-redshift quasar, two spectra might be required:
	the first to confirm the candidate, and a second, higher \sn spectrum, to measure the damping wing.
	As a fiducial value, we adopt the requirement that confirmation that a source is a quasar, even if weak lined, 
	and measurement of a redshift, requires an observed S/N $\gtrsim1.2$ per \AA, 
	in the continuum, over a wide wavelength range\footnote{
	A \sn per pixel of \(1.13/\AA\) based on measuring the flux over \(100\,\AA\) 
	redwards of a possible \lya break would result in a reasonable detection of the break (\(\sn=8\)).}.
	A spectrum for measuring the damping wing would require S/N $\gtrsim4$ per \AA, 
	or an integration time ten times longer than required for identification\footnote{
	The spectrum of \uj presented by \citet{Mortlock2011} has a \sn per \AA~slightly above 4,
	and was deep enough to place a lower limit on the neutral hydrogen fraction.
	We therefore consider this to be an appropriate lower limit on the depth of a spectrum suitable for measuring the damping wing.}. 
	
	Using these numbers and allowing for a maximum of 3\,h integration time to classify any candidate, 
	a campaign of spectroscopic confirmation of \euc sources down to $J=22$ with 8\,m class telescopes would be feasible. 
	Some sources with strong lines that are fainter than $J=22$ could be identified, 
	but we are mainly interested in creating a complete sample, 
	in order to measure the luminosity function. 
	This calculation is somewhat conservative. 
	It is based on data taken with the Gemini GNIRS instrument, 
	assuming mediocre seeing conditions (up to \(1\arcsecond\)), 
	as might be appropriate as a specification for a large programme. 
	It would be possible to reach deeper in better seeing (see, e.g., the results achieved by \citealt{Kriek2015}).
	Allowing for 10\,h observing time per source to measure the damping wing, 
	a campaign of spectroscopy to measure the cosmic neutral fraction as a function of redshift, 
	and its variance, could reach $J=21.3$ with 8\,m class telescopes. 
	These two limits are marked by lines in Fig. \ref{fig:Mzplane}.

	This analysis illustrates the difficulty of completing 
	all the potential high-redshift quasar science with 8m class telescopes. 
	To confirm sources fainter than $J\sim 22$, 
	and to measure the faint end of the quasar luminosity function 
	will require future facilities such as the \textit{James Webb} Space Telescope (JWST) 
	or the European Extremely Large Telescope.
	At very faint magnitudes (\(J\sim23\)),
	where the number of contaminants may be comparable to the number of quasars (Sect.~\ref{sec:contam}), 
	it seems likely that a spectroscopic campaign would need to be limited to a subset of the candidates. 
	The limits presented in this discussion are illustrative only, and
	the approach to follow-up will naturally depend on
	the details of future candidate lists, and the availability of follow-up resources. 
	
    It will be possible to identify some of the brightest candidates using the slitless spectra obtained with the \euc NISP
    instrument, but the wavelength coverage of the instrument is no longer very well suited to this task.
	NISP will produce slitless (R\,=\,380) spectra of all sources in the wide survey.
	The NISP red grism originally covered the wavelength range  (\(\myrange{1.1}{2.0}\,\micron\)) \citep{Laureijs2011}, 
	and for this configuration \citet{Roche2012} showed how high-redshift quasars $z>8$ 
	could be identified by the continuum break at \lya. 
	In the new configuration the wavelength coverage is (\(\myrange{1.25}{1.85}\,\micron\)), 
	and the \lya break is outside this range for all redshifts of interest $7<z<9$. 
	Quasars in this redshift range might still be identified 
	by the detection of \Civ1549 and \(\left.{\rm C}\,\textsc{iii}\right]\) 1909.
	To explore this approach further, we ran exploratory simulations 
	using an early version of the \euc simulator of the NISP, 
	named TIPS \citep[TIPS Is a Pixel Simulator;][]{Zoubian2014}. 
	These simulations consisted in producing 2D grism images of the observed spectrum of ULAS J1120+0641
	\citep[][\(J=20.2\)]{Mortlock2011},
    and did not include source contamination. 
    Simple image stacks and visual inspection suggested that sources with \(J < 21\) 
    could be identified by this means.

\subsection{Choice of contaminating populations}
\label{sec:cosmosdiscuss}

\begin{figure}[t]
	\centering
	\includegraphics[width=6cm,trim={0cm 0cm 0cm 0cm}]{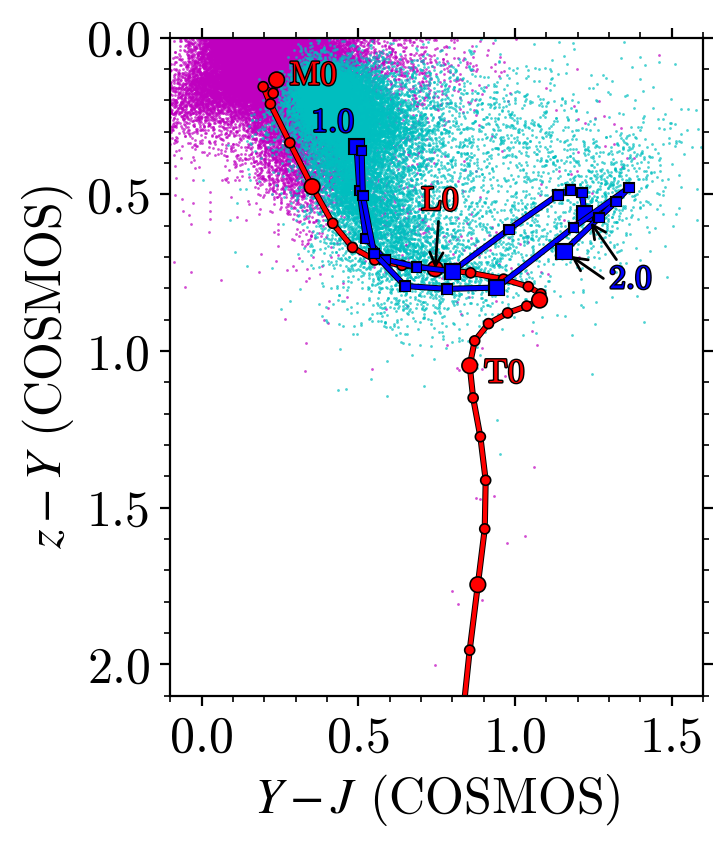}
	\includegraphics[width=6cm,trim={0.0cm 0cm 0cm 0cm}]{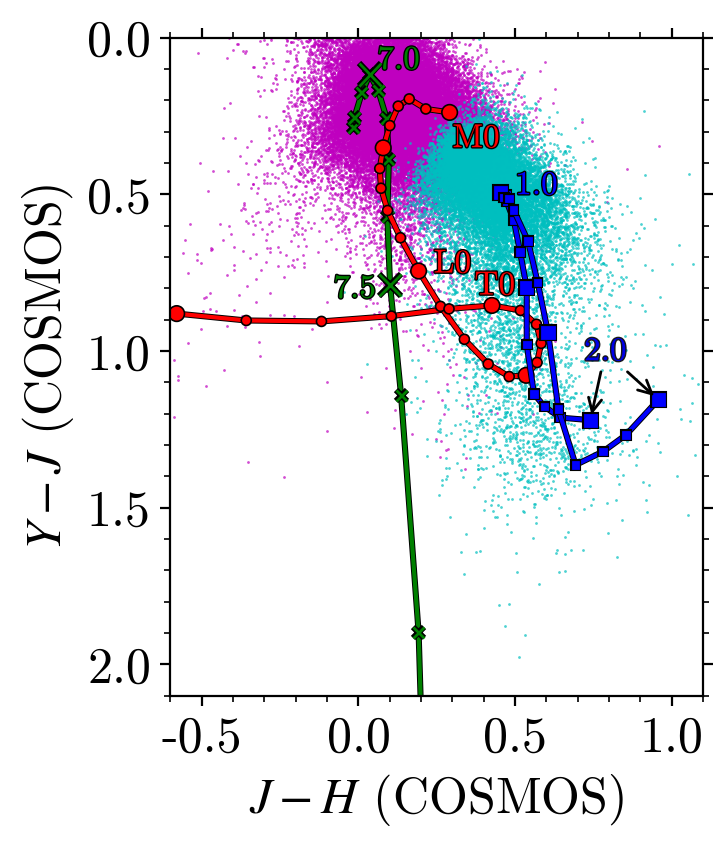}
	\caption{Model COSMOS colour tracks and COSMOS sources.
			The COSMOS filters are different to \euc and LSST/Pan-STARRS,
			resulting in slight differences in the tracks presented in Fig.~\ref{fig:colourcolour}.
			The individual population models are however indicated in the same way as in that figure:
			the red tracks show MLTs; the early-type elliptical tracks are blue; and the quasar track is green.
			We additionally plot sources brighter than \(J=23\) in the COSMOS catalogue \citep{Laigle2016}.
			Magenta points indicate sources with photometry best fit by an MLT model.
			The cyan points indicate sources with photometry best fit by an elliptical galaxy model.
			\textit{Upper:} \(zYJ\) colours. \textit{Lower:} \(YJH\) colours.}
	\label{fig:cosmoscolours}
\end{figure}

	The predictions in this work are based on the assumption that the relevant population space
	for high-redshift quasar searches can be reduced to the target quasars, and two types of contaminants:
	namely, MLTs and elliptical galaxies.
	These populations have long been known as important sources of contamination 
	for \(z>7\) quasar searches \citep[e.g.][]{Hewett2006}.
	MLTs and ellipticals are expected to be abundant in \euc, and as seen in Fig.~\ref{fig:colourcolour}, 
	their near-infrared colours match closely to those of quasars.
	One might ask, however, whether this fully represents the range 
	of populations present in real data
	that might have a bearing on quasar selection.
	
	To investigate this question we return to the COSMOS sample presented by \citet{Laigle2016},
		which we used in Sect.~\ref{sec:mg} to model the elliptical surface density.
		We determine model \(zYJH\) COSMOS colours from our population templates,
		and show these tracks in Fig.~\ref{fig:cosmoscolours},
		along with the 77\,000 sources brighter than \(J=23\) in the COSMOS catalogue.
    Fig.~\ref{fig:cosmoscolours} indicates that our choice of contaminant templates
    encapsulates the red envelope of the COSMOS sources very well.
    We do not see evidence for a significant additional population
    that is very red in \(z-Y\) (where the reddest sources follow the T-dwarf track closely), or in \(Y-J\),
    although in the latter case, a few sources appear to scatter towards \(Y-J\sim2\),
    consistent with \(z>7.7\) quasars.    

    As a further check we apply our minimum-\(\chisq\) selection method to the COSMOS sources.
    Despite the depth of the COSMOS survey, the small area (\(<1.5\,\dsq\)) 
    means we do not expect any \(z>7\) quasars to be present in the catalogue.
    Therefore any significant number of candidates that are better fit by a quasar template
    would likely be indicative of an additional population that needs to be accounted for in our selection methods.
    However, on the basis of our SED fitting, 
    all sources are classified as either an MLT or as an elliptical galaxy,
    respectively indicated in magenta and cyan in Fig.~\ref{fig:cosmoscolours}.
    This result lends support to the above statement 
    that there is no significant additional contaminating population in the COSMOS data.
    We note our ability to distinguish quasars from contaminants 
    is in this case helped by very deep COSMOS \(z\)-band data.
    Nevertheless, in general terms these preliminary results indicate
    that our models are representative of the populations that will be of concern once \euc data are available.

\subsection{Importance of the early-type galaxy population}
\label{sec:elldiscuss}

\begin{figure}[t]
	\centering
	\includegraphics[width=9cm]{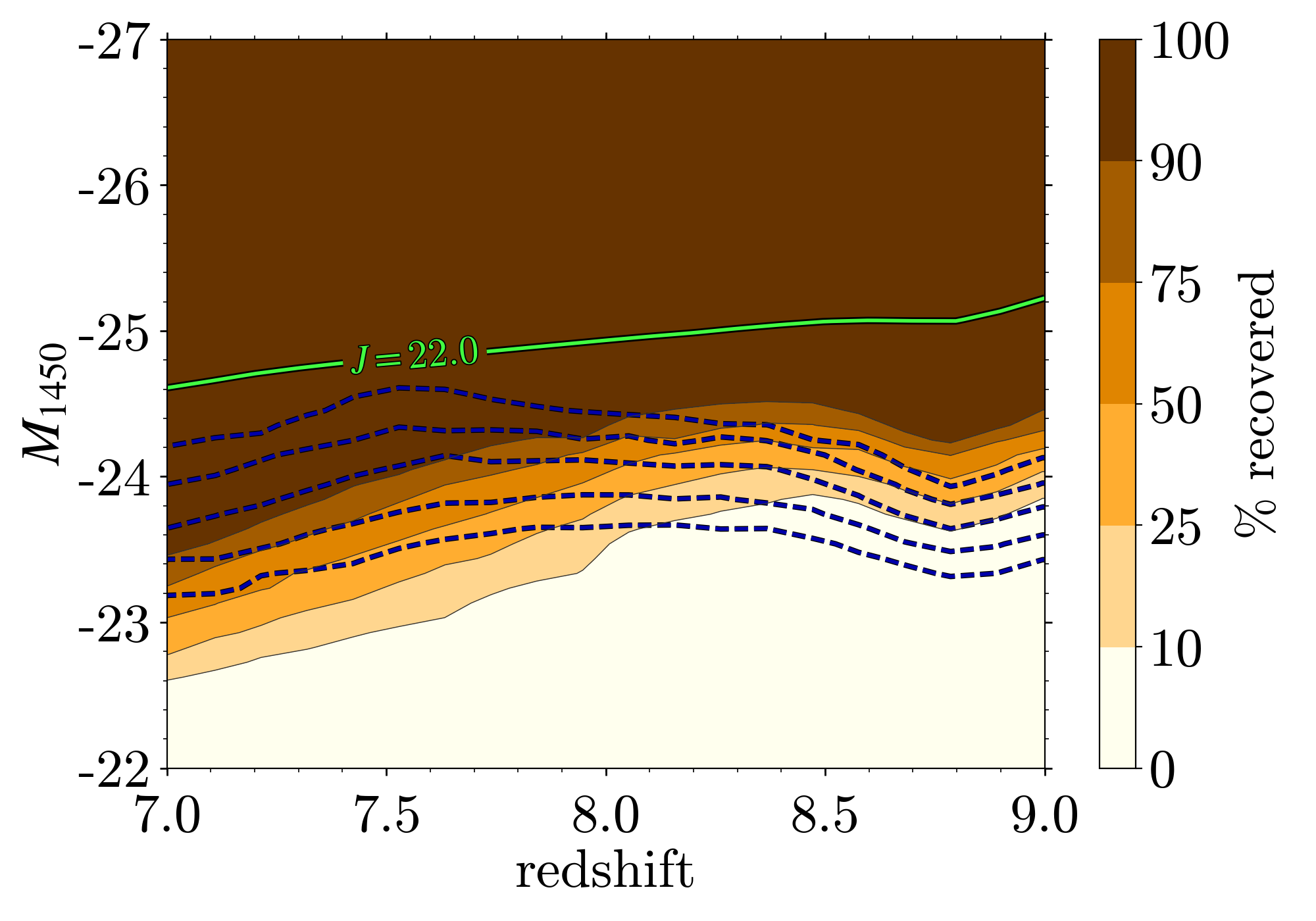}
	\caption{Quasar selection functions using ground-based data
		and using BMC, assuming a single contaminating population.
		Filled contours indicate the case where only galaxies 
		are considered as contaminants, i.e., \(W_\mathrm{s} = 0\).
		Dashed lines indicate the case where only MLTs
		are considered, i.e., \(W_\mathrm{g} = 0, W_\mathrm{s} \neq 0\).
		Contour intervals are the same in both cases.
		The contour indicating \(J=22\) is also shown as a solid green line.
	}
	\label{fig:sepmltell}
\end{figure}	

In Sect.~\ref{sec:mg} close attention was paid to the sizes 
of early-type galaxies in the redshift range of interest $1<z<2$.
We were motivated by the concern that faint (\(J>22\)) galaxies may be mistaken for point sources,
given the relatively large pixel size of the \euc NISP instrument, 
and the small half-light radii of the galaxies at faint magnitudes. 
Based on the predicted sizes we conservatively assumed 
that all faint $J>22$ early-type galaxies would be classified as point sources. 
We now consider whether this assumption is significant in terms of the predicted quasar numbers.

To proceed, we produce two further selection functions,
for the case where ground-based \(z\)-band data are available, i.e., to be compared with Fig.~\ref{fig:qsfg}.
In each case we switch off the effect of either the MLT or elliptical population.
That is to say, we set either \(W_\mathrm{s}\) or \(W_\mathrm{g}\) equal to zero when calculating \(P_{\mathrm q}\). 
We present the resulting two selection functions in Fig.~\ref{fig:sepmltell}.
In this plot the MLT selection function (i.e., where MLTs are the only contaminating population) 
is shown by the dashed lines and the ellipticals selection function as the solid lines. 
At any redshift the population that dominates the contamination, 
and controls the quasar detection probability, 
is the population where the contours are higher up the plot (towards brighter magnitudes). 
Since the assumptions we have made about the early-type galaxies have been conservative, 
the true contours for this population are somewhat lower down the plot.

At \(z<8\), the MLT population dominates the selection function,
i.e., the assumptions about the early-type galaxies have no influence on the predicted quasar numbers over the MLTs. 
At higher redshifts, the situation changes.
At $z\sim8$ the contours cross, meaning the two populations contribute approximately equally, 
while at $z>8.2$ the overall selection function is constrained 
more tightly by the ellipticals, but not by much. 
This indicates that even if it is possible to eliminate 
most of the early-type galaxies within the \euc pipeline on the basis of morphology, 
the improvement in the number of quasars detectable will be fairly modest.
To test this, we integrated the QLF over the dashed selection function 
in Fig.~\ref{fig:sepmltell} (with \(W_\mathrm{g}=0\)),
to evaluate the predicted quasar numbers with MLTs as the sole contaminant.
Compared to Table~\ref{tab:counts}, we find a negligible change for \(z<8\),
while the predicted counts at \(z>8\) increase by \(25\%\).

\subsection{Variations within the quasar population}
\label{sec:lx_cx}

\begin{table}[t] 
	\centering
	\advance\leftskip-3cm
	\advance\rightskip-3cm
	\caption{Weights for each value of line width and continuum slope,
	used to combine the individual \euc quasar selection functions.
    The total weight of a single quasar model is given by the product of the relevant slope and line width.}
	\label{tab:modelweights}
	\begin{tabular}{ccc}
	  \hline \hline \\[-2ex]
	  parameter & model value & weight \\ 
	  \hline \\[-2ex]
      line width & half   &  0.3 \\
                 &standard & 0.6   \\
                 &double & 0.1  \\
      \hline \\[-2ex]
	  continuum slope &blue & 0.05     \\
	                  &standard & 0.7  \\
	                  &red & 0.25      \\
	  \hline		
\end{tabular}

\end{table}

In simulations of the quasar selection function so far
we have made a simplifying assumption by only using a single typical model spectrum to generate synthetic quasar colours. 
We have not considered variations in the continuum and line emission.
As explained in Sect.~\ref{sec:mq}, the reason for using a single model quasar is for simplicity, 
because this still yields accurate estimates of the quasar yield (e.g., Barnett et al. in prep.). 
In this subsection we explore the sensitivity of the selection function calculation 
to a mismatch between the actual quasar colours, and the model used in the selection. 
To proceed, we model quasars with a range of spectral types,
and consider selection using a single, typical type.
In the actual search we will use an appropriate range of model spectra in the selection, 
so the calculation presented here will overestimate the effect of spectral mismatch. 
Nevertheless it gives a sense of the scale of the problem of a mismatch between the actual quasar SEDs 
and those assumed in the selection algorithm, 
and so gives an indication of the proportion of quasars that might be missed 
if the spectra of quasars at $z>7$ are more diverse than at lower redshifts.

The \citet{Hewett2006} and \citet{Maddox2008} quasar models are available
for a range of continuum slopes and emission line strengths.
We now wish to produce selection functions for quasar populations with different properties,
and compare the resulting numbers with those presented in Sect.~\ref{sec:qsf}.
Explicitly, the typical quasar spectrum that we have used so far 
has a line strength with rest frame \(\mathrm{EW_{\Civ} = 39.1\,\AA}\)
and UV continuum slope \(f_{1315}/f_{2225}=cs=1.0\).
We now additionally make use of models with doubled and halved line strengths,
and blue and red continuum slopes
corresponding to \(cs=1.16\), and 0.84 respectively. 
Broad absorption line quasars are effectively included here, 
as the colours would be matched by weak-lined objects with red continua.
We therefore have nine combinations of line widths and slopes in total.
Following the previous prescription we simulate grids of quasars for each type,
using the \euc \(O\) band in the optical.
We then determine selection functions and combine them,
with the weight for each value of line width and continuum slope given in Table~\ref{tab:modelweights}.
These weights are based on the distribution of slopes and line widths 
measured for SDSS DR7 quasars, i.e., representative of the quasar population at redshifts $0<z<6$.
Conceivably the distributions of line widths and continuum slopes 
in the quasar population at $z>7$ may prove to be somewhat different  
(see, e.g., the evidence for a larger fraction of sources with weak \Civ emission
	at \(z>6\) presented by \citealt{Shen2019}),
in which case the relative weights in the selection, 
or indeed the models themselves, can be adjusted.

Integrating the \citet{Jiang2016} QLF ($k=-0.72$) over the resulting selection function,
there is a clear impact on the predicted numbers 
	when there is a mismatch between the quasar template(s) used in the selection method,
	and the actual quasar SEDs. 
The total yield over \(7<z<9\) is reduced by \(20\%\),
compared to the total of 124 sources predicted previously
(Table~\ref{tab:counts}, column 1).
This decrease is driven by slightly worse selection at \(z<8\).

	In contrast, preliminary simulations suggest that
	incorporating the full range of quasar templates in the BMC method
	mitigates substantially against this reduction in numbers, as stated previously.
	In this case, we draw 1000 quasars distributed uniformly in redshift and luminosity space
	from each of the nine grids described above.
	We then determine \pq values for these quasars using all nine SEDs in the BMC,
	weighting the contribution of each SED to \(W_\mathrm{q}\)
	using the values in Table~\ref{tab:modelweights}.
	We find the number of recovered quasars is almost identical
	to the case where we use just the typical model 
	both to generate synthetic quasars and in the BMC (i.e., as in the rest of the paper).
	The results from this section therefore highlight the need 
	for a realistic range of model templates in our selection methods,
	in order to maximise the future quasar yield with \euc.



\section{Summary}
\label{sec:sum}

In this paper we have presented a detailed study of the use of the
\(15\,000\,\dsq\) \euc wide survey for the discovery of quasars in
the redshift range \(7<z<9\), updating the predicted quasar yield
presented by \citet{Laureijs2011}. 
This work incorporates revisions to the NISP filter wavelengths and the planned survey area, 
and accounts for the steeper redshift evolution of the
quasar number density, 
based on the decline measured over
\(z=\myrange{5}{6}\) by \citet{Jiang2016}. 
We have extended the \citet{Laureijs2011} analysis, that considered redshifts \(8<z<9\),
to include the range \(7<z<8\), 
and we have improved the earlier study in two important ways: 
candidate quasars are now selected using statistical methods rather than heuristic colour cuts, 
allowing the detection of fainter high-redshift quasars; 
and we have developed more accurate models of the contaminant populations,
i.e., MLT dwarfs, and early-type galaxies at redshifts $1<z<2$.

The main results of this paper are based on simulations 
of \euc quasar selection functions and contaminating populations, and are summarised below.
\begin{enumerate}
	
	\item Quasars with redshifts $8<z<9$ can be selected from \euc data alone. 
	Even if the rate of decline in the space density of quasars accelerates beyond $z=6$, 
	and is as steep as $\Phi\propto10^{k(z-6)}$, $k=-0.92$, 
	there should be some 6 quasars discoverable with \euc at $z>8$, brighter than $J\sim23$,
	using \euc data alone, improving to 8 quasars if deep
	ground based data is available.
	
	\item Deep ground-based \(z\)-band data from LSST and Pan-STARRS 
	significantly boost the selection of quasars over \(7<z<8\), 
	compared to using the \euc $O$ optical band, 
	due to the sharper contrast for a spectral break to the blue of the $Y$ filter. 
	Using the expected depths for the optical surveys, 
	we find that \euc will discover more than 100 quasars $7<z<8$, assuming $k=-0.92$. 
	If \(z\)-band data are not available, 
	the total return is smaller by a factor greater than two.
	
	\item Both the BMC and minimum-\(\chisq\) method are able to eliminate the majority of contaminants,
	although at lower \sn contamination from MLTs and ellipticals needs to be considered 
	and may impact future follow-up strategy.
	Over the redshift range \(7<z<8\), the inclusion of priors means the BMC method can reach at least
	0.5\,mag fainter than the simpler minimum-\(\chisq\) method,
    resulting in a factor of two difference in the total predicted numbers.
		
	\item The rate of decline of the QLF, 
	parametrised by $k$, is the most significant unknown for the number count predictions. 
	If $k=-0.72$ over the redshift range $7<z<8$ this parameter 
	will be measured to a $1\sigma$ uncertainty of 0.07.
	
\end{enumerate}

We anticipate that, except for the brightest sources, spectroscopic follow-up of \euc quasar candidates
will generally be challenging with existing ground-based 8\,m telescopes. 
Nevertheless, beginning with \euc Data Release 1, planned for 2024, 
we expect significant numbers of \(z>7\) quasars to be discovered with \euc,
allowing detailed studies of the cosmic neutral fraction of hydrogen over redshifts $7<z<9$, 
which will make an important contribution towards understanding the process of cosmic reionisation. 
These new samples will also be valuable for studies of early  
SMBH growth, from the measurement of black hole masses in individual sources and through additional constraints on the faint end of the QLF. Conceivably for some sources it may prove possible to image the IGM structure surrounding the quasar in the light of \lya.

\begin{acknowledgements}
	We would like to acknowledge the anonymous A\&A referee,
	who provided useful and detailed comments on this manuscript.
	The \euc internal referees were Francisco Castander and Johan Fynbo, 
	whose reports led to several improvements in the presentation of this paper.
	We are also grateful to several members of the \euc Primeval Universe Science Working Group, 
	who provided comprehensive comments and suggestions for this work, 
	which have led to significant changes in the analysis and discussion:
	Rebecca Bowler; Peter Capak; Pratika Dayal;
	Andrea Ferrara; Kari Helgason; and Michael Strauss.
	We would additionally like to thank Konrad Kuijken and Jean-Charles Cuillandre for their advice on Pan-STARRS.
	\AckEC
	This work was supported by grant ST/N000838/1 
	from the Science and Technology Facilities Council. 
	This research has benefitted from the SpeX Prism Spectral Libraries, 
	maintained by Adam Burgasser at http://pono.ucsd.edu/~adam/browndwarfs/spexprism.
\end{acknowledgements}

\bibliography{euclidrefs_new}{} \bibliographystyle{aa}

\end{document}